\def\simlt{\lower.5ex\hbox{$\; \buildrel < \over \sim \;$}}

\RequirePackage{lineno} 
\documentclass[preprint2]{emulateapj}
\usepackage[all,cmtip]{xy}
\usepackage{amsmath}
\usepackage[subnum]{cases}
\usepackage{mathtools}
\usepackage[english]{babel}
\usepackage{blindtext}
\usepackage{listings}
\usepackage{hyperref}
\allowdisplaybreaks
\newcommand*{\rom}[1]{\expandafter\@slowromancap\romannumeral #1@}
\newcommand{\myemail}{xiz@ucsc.edu}
\slugcomment{Accepted at ApJ}
\shorttitle{Vertical Tracer Mixing in Planetary Atmospheres Part II}
\shortauthors{Zhang \& Showman}
\begin{document}

\title{Global-mean Vertical Tracer Mixing in Planetary Atmospheres \\
II: Tidally Locked Planets}
\author{Xi Zhang$^1$ and Adam P. Showman$^2$}
\affil{$^1$Department of Earth and Planetary Sciences, University of California Santa Cruz, CA 95064}
\affil{$^2$Department of Planetary Sciences and Lunar and Planetary Laboratory, University of Arizona, AZ 85721}
\altaffiltext{1}{Correspondence to be directed to \myemail}

\begin{abstract}

In Zhang $\&$ Showman (2018, hereafter Paper I), we developed an analytical theory of 1D eddy diffusivity $K_{zz}$ for global-mean vertical tracer transport in a 3D atmosphere. We also presented 2D numerical simulations on fast-rotating planets to validate our theory. On a slowly rotating planet such as Venus or a tidally locked planet (not necessarily a slow-rotator) such as a hot Jupiter, the tracer distribution could exhibit significant longitudinal inhomogeneity and tracer transport is intrinsically 3D. Here we study the global-mean vertical tracer transport on tidally locked planets using 3D tracer-transport simulations. We find that our analytical $K_{zz}$ theory in Paper I is validated on tidally locked planets over a wide parameter space. $K_{zz}$ strongly depends on the large-scale circulation strength, horizontal mixing due to eddies and waves and local tracer sources and sinks due to chemistry and microphysics. As our analytical theory predicted, $K_{zz}$ on tidally locked planets also exhibit three regimes In Regime I where the chemical and microphysical processes are uniformly distributed across the globe, different chemical species should be transported via different eddy diffusivity. In Regime II where the chemical and microphysical processes are non-uniform---for example, photochemistry or cloud formation that exhibits strong day-night contrast---the global-mean vertical tracer mixing does not always behave diffusively. In the third regime where the tracer is long-lived, non-diffusive effects are significant. Using species-dependent eddy diffusivity, we provide a new analytical theory of the dynamical quench points for disequilibrium tracers on tidally locked planets from first principles.

\end{abstract}

\keywords{planetary atmospheres - eddy mixing - tracer transport - methods: analytical and numerical}

\section{Introduction}

As stated in Paper I (\citealt{zhang2018kzz}), atmospheric transport can drive the atmospheric chemical species out of chemical equilibrium and greatly influence the observations (e.g., \citealt{prinn1976chemistry}, \citealt{prinn1977carbon}, \citealt{smith1998estimation}, \citealt{cooper-showman-2006}, \citealt{visscher-moses-2011}). Currently, the dominant approach of simulating the global-mean vertical distributions of chemical species and clouds on atmospheres of tidally locked planets is solving a 1D diffusive system with an effective eddy diffusivity and chemical/microphysical source and sinks (e.g., \citealt{moses-etal-2011}, \citealt{line2011thermochemical}, \citealt{tsai2017vulcan}, \citealt{helling2008dust}, \citealt{gao2014bimodal}, \citealt{lavvas2017aerosol}, \citealt{powell2018formation}). In the 1D chemical-diffusion framework, the strength of the vertical diffusion is characterized by a parameter called eddy diffusivity or eddy mixing coefficient $K_{zz}$. The physical underpinning of this key parameter is obscure. In Paper I, we constructed a first-principles theory of $K_{zz}$ and provided analytical expression for $K_{zz}$ in a 3D atmosphere under idealized assumptions about the nature of the chemical source/sink and other simplifications. We further validated the theory using 2D numerical simulations on fast-rotating planets. However, for tidally locked exoplanets or slow-rotating planets that exhibit significant day-night contrast such as Venus, a 2D model is not sufficient to simulate the intrinsically 3D tracer transport process in those planetary atmospheres. This is our focus in this paper.

On these planets, the global circulation pattern could vary from a substellar-to-anti-stellar circulation in the upper atmosphere to a fast zonal-jet circulation in the lower atmosphere (e.g., \citealt{bougher1997venus} for Venus, \citealt{showman-etal-2013} and  \citealt{cooper-showman-2005} and \citealt{zhang2017effects} for tidally locked planets). Tracer transport with a substellar-to-anti-stellar wind pattern cannot be treated in a 2D zonal-symmetric system as we described in Paper I. Even in the zonal-jet regime, the existence of longitudinally varying dynamical structures (eddies) plays a critical role in both the dynamics and the mixing of chemical tracers. Moreover, the day-night temperature and insolation gradients imply that, for many tracers, there will exist large day-night variations in chemical sources and sinks that further exacerbate the longitudinal asymmetry and amplify the overall 3D nature of the problem. For instance, photochemically produced tracers on a slowly rotating planet or a tidally locked planet could develop substantially different chemistry and equilibrium tracer distributions between the dayside and nightside, a canonical example being sulfur species from photochemistry on Venus (e.g., \citealt{zhang2010venus}, \citealt{zhang2012sulfur}). For another example, transport of haze and cloud particles on a tidally locked planet is essentially 3D because their formation and destruction are greatly affected by the local temperature and condensable gas abundance which might have a large zonal variation (\citealt{powell2018formation}). To investigate the global-mean vertical tracer transport on tidally locked exoplanets, a 3D dynamical model is needed (e.g., \citealt{cooper-showman-2006}, \citealt{parmentier20133d}).

There have been several studies of 3D chemical-transport on tidally locked exoplanets with simplified chemical and cloud schemes (e.g., \citealt{cooper-showman-2006}, \citealt{parmentier20133d}, \citealt{charnay20153d2}, \citealt{lee2016dynamic}, \citealt{drummond2018effect}, \citealt{lines2018simulating}). However, how to parameterize the the 3D transport processes in a 1D global-mean tracer transport model is still an open question. Especially, if one adopts the 1D chemical-diffusion framework, a thorough theoretical link between the 3D tracer transport and the effective 1D eddy diffusivity $K_{zz}$ on these planets has not been well established. For example, \citet{parmentier20133d} studied simple passive cloud tracers using a 3D dynamical model, and showed that the 1D effective eddy diffusivity on hot Jupiters does not exhibit a significant dependance on tracer sinks. This seems contradictory to our findings as well as previous terrestrial studies such as \citet{holton1986dynamically} that the $K_{zz}$ should be species-dependent. This puzzle has not been satisfactorily solved. 

Here we will apply our analytical $K_{zz}$ theory in Paper I and 3D tracer-transport simulations to the photospheres on tidally locked planets where most of the chemical tracers and haze/clouds are observed. In the following sections, we will first recap the analytical $K_{zz}$ theory in Paper I and further develop it for tidally locked planets. A 3D general circulation model (GCM) with passive tracers is then used to understand the atmospheres on tidally locked exoplanets under two typical scenarios: tracers that are transported from the deep atmosphere and that are produced in the upper atmospheres via photochemistry or ion chemistry. We will also study the effects of our $K_{zz}$ theory on the dynamical quenching of disequilibrium tracers on tidally locked planets. We conclude this study with several take-home messages.

\section{$K_{zz}$ Theory for Tidally Locked Planets}

\subsection{General Theory of $K_{zz}$}

Our $K_{zz}$ theory in Paper I is based on a linear chemical scheme which relaxes the tracer mixing ratio $\chi$ towards local chemical equilibrium $\chi_0$ over a timescale $\tau_c$. The chemical source/sink term is thus written as $(\chi_0-\chi)/\tau_{c}$. We also defined $\chi_0=\overline{\chi_0}+\chi_0^{\prime}$, where $\overline{\chi_0}$ is the global-mean of $\chi_0$ which is only a function of pressure, and $\chi_0^{\prime}$ is the departure of equilibrium tracer abundance from its global mean. We have demarcated three atmospheric regimes of $K_{zz}$ in terms of the tracer chemical lifetime and horizontal tracer distribution under chemical equilibrium\footnote{Here, as in Paper I, we use the term ``chemistry" to represent any non-dynamical processes that affect the tracer distribution, such as chemical reactions in the gas and particle phase, haze and cloud formation, or other phase transition processes.}, and derived analytical formulae of $K_{zz}$ for each regime. 

Regime I is for a short-lived tracer with chemical equilibrium abundance uniformly distributed across the globe. In this regime, the global-mean tracer mixing behaves diffusively and $K_{zz}$ can be expressed as:
\begin{equation}
K_{zz}\approx\frac{\overline{w^2}}{\tau_{d}^{-1}+\tau_{c}^{-1}}\approx\frac{\hat{w}^2}{\hat{w}L_v^{-1}+\tau_{c}^{-1}}.
\end{equation}
Here $w$ is the vertical velocity and $\hat{w}=(\overline{w^2})^{1/2}$ is the root-mean-square of $w$. The horizontal tracer mixing timescale $\tau_{d}\approx L_h/U_{adv}$ where $L_h$ is horizontal characteristic length scale and $U_{adv}$ is the horizontal wind speed. As pointed out by \citet{showman-polvani-2011} and \citet{perez-becker-showman-2013}, the horizontal scales of dominant flow patterns on typical hot Jupiters are comparable to the Rossby deformation radius, which is similar to the planetary radius $a$. Thus we assume $L_h\sim a$ in this study. From the continuity equation, $\tau_d$ can also be estimated from the vertical dynamical timescale $\tau_d=L_h/U\approx L_v/\hat{w}$ where $L_v$ is the vertical transport length scale (\citealt{komacek2016atmospheric}). 

Regime II is for a short-lived tracer with a non-uniform distribution of the chemical equilibrium abundance. In this regime, the non-diffusive behavior in the global-mean tracer mixing could be important. It can be represented by an additional term for $K_{zz}$:

\begin{equation}
  \begin{aligned}
K_{zz}&\approx\frac{\overline{w^2}}{\tau_{d}^{-1}+\tau_{c}^{-1}}-\frac{\overline{w\chi_0^{\prime}}}{1+\tau_{d}^{-1}\tau_{c}}(\frac{\partial\overline{\chi}}{\partial z})^{-1} \\
&\approx\frac{\hat{w}^2}{\hat{w}L_v^{-1}+\tau_{c}^{-1}}-\frac{\hat{w}\Delta\chi_0^{\prime}}{1+\hat{w}L_v^{-1}\tau_{c}}(\frac{\partial\overline{\chi}}{\partial z})^{-1},
 \end{aligned}
\end{equation}
where $\Delta\chi_0^{\prime}$ is the magnitude of $\chi_0^{\prime}$. The second term on the right hand side depends on the vertical gradient of the global-mean tracer profile and deviation of the tracer mixing ratio from its global mean. 

Regime III is a long-lived tracer regime with the tracer material surface significantly controlled by dynamics. In fact, no simple analytical theory of the global-mean vertical tracer transport exists in this regime due to the complicated dynamical transport behavior. We introduced an analytical formula of $K_{zz}$ by letting $\tau_c\rightarrow \infty$ in Eq. (1) or (2):

\begin{equation}
K_{zz}\approx\overline{w^2}\tau_{d}=\hat{w}L_v.
\end{equation}

Note that the above three expressions can be unified in a single expression in Eq. (2). Taking $\chi_0^{\prime}$ for the uniform $\chi_0$ case, Eq. (2) will be reduced to Eq. (1). Taking $\tau_c\rightarrow \infty$ for very long-lived tracers, Eq. (2) will be reduced to Eq. (3). 

\subsection{Vertical Velocity Scaling on Tidally Locked Planets}

In order to fully predict $K_{zz}$ from first principles, we also need to analytically estimate the vertical velocity scale $\hat{w}$. For tidally locked giant planets, $\hat{w}$ can be estimated from first principles using the scaling theory developed in \citet{komacek2016atmospheric} and \citet{zhang2017effects}. First, the horizontal velocity scale $U$ can approximated as (see Appendix A in \citealt{zhang2017effects}):
\begin{equation}
\frac{U}{U_{eq}}\sim\sqrt{(\alpha/2\gamma)^2+1}-\alpha/2\gamma
\end{equation}
where the non-dimensional parameters $\alpha$ and $\gamma$ are defined as:
\begin{align}
\alpha&=1+\frac{(\Omega+{\tau_{drag}^{-1}})\tau_{wave}^2}{\tau_{rad}\Delta\ln p}
\\
\gamma&=\frac{\tau_{wave}^2}{\tau_{rad}\tau_{adv,eq}\Delta\ln p}.
\end{align}

Here $\tau_{wave}\sim L/NH$ is the characteristic wave propagation timescale over a typical horizontal dynamics length scale $L$ and $N$ is the buoyancy frequency (Brunt-V$\ddot{\mathrm{a}}$is$\ddot{\mathrm{a}}$l$\ddot{\mathrm{a}}$ frequency) of the atmosphere. $\tau_{adv,eq}\sim L/U_{eq}$ is the advective timescale due to the ``equilibrium cyclostrophic wind'' $U_{eq} = (R\Delta T_{eq}\Delta\ln p/2)^{1/2}$. 

As noted, we assume the typical horizontal length scale $L\sim a$. Note that the continuity equation states $U/a\sim \hat{w}/H$. The vertical velocity scale $\hat{w}$ can then be estimated as:
\begin{equation}
\hat{w}\sim \frac{U_{eq}a}{H}(\sqrt{(\alpha/2\gamma)^2+1}-\alpha/2\gamma).
\end{equation}

Insert Eq. (7) into Eq. (1-3) and we obtain the analytical expression of $K_{zz}$ for tidally locked exoplanets from first principles. Next, we will perform 3D numerical experiments to quantitatively verify our theoretical arguments.

\section{3D Simulation on Tidally Locked Planets}

The observable atmosphere of most close-in tidally locked planets such as hot Jupiters is stratified due to the strong irradiation from their host stars (e.g., \citealt{fortney2008unified}, \citealt{madhusudhan2009temperature}, \citealt{line2012information}). The atmospheric dynamics is approximately governed by the hydrostatic primitive equations (e.g., \citealt{showman-etal-2009}). In pressure ($p$) coordinates, the equations including tracer transport are:
 \begin{subequations}
 \begin{align}
\frac{D\vec{\bf{u}}}{Dt}+f\hat{\bf{k}}\times \vec{\bf{u}}+\nabla_p\Phi&=F_{d}
\\
\frac{\partial{\Phi}}{\partial{p}}&=-\frac{1}{\rho}
\\
\nabla_p\cdot \vec{\bf{u}}+\frac{\partial{\omega}}{\partial{p}}&=0
\\
\frac{DT}{Dt}-\frac{\omega}{\rho c_p}&=\frac{q}{c_{p}}
\\
p&=\rho RT
\\
\frac{\partial \chi}{\partial t}+\vec{\bf{u}}\cdot\nabla_p\chi+\omega\frac{\partial \chi}{\partial p}&=S
\end{align}
\end{subequations}
where $\rho$ is density; $T$ is temperature; $c_p$ is the specific heat at constant pressure in units of $\mathrm{J~Kg^{-1}~K^{-1}}$; $R$ is gas constant in units of $\mathrm{J~Kg^{-1}~K^{-1}}$; $\Phi=gz$ is geopotential where $g$ is gravitational acceleration assumed constant in our study; $z$ is altitude; $f=2\Omega \sin{\phi}$ is Coriolis parameter where $\Omega$ is the planetary rotation rate and $\phi$ is latitude; $\nabla_p$ is the horizontal gradient at constant pressure; $\omega=Dp/Dt$ is vertical velocity in pressure coordinate that can be converted to vertical velocity in the log-pressure coordinate $w$; $\vec{\bf{u}}=(u, v)$ is the horizontal velocity at constant pressure where $u$ is the zonal (east-west) velocity and $v$ is the meridional (north-south) velocity; $q$ is the radiative heating/cooling rate. $D/Dt=\partial/\partial{t} + \vec{\bf{u}}\cdot\nabla_p+\omega\partial/\partial{p}$ is the material derivative; $F_{d}$ is a drag term representing the missing physics such as magnetohydrodynamics (MHD) drag or sub-grid turbulent mixing (e.g., \citealt{li-goodman-2010}; \citealt{youdin2010mechanical}). $\chi$ is the mixing ratio of a tracer with chemical source/sink $S$.

\begin{deluxetable*}{ccccccc}[bp]
\centering
\tablecolumns{3} 
\tablewidth{0pt}  
\tablecaption{3D Tracer Chemical-dynamical Simulation cases.}
\tablehead{\colhead {Experiment}			&
\colhead {Tracer}  &
\colhead {$\chi_0$} &
\colhead {Chemical Timescale} 
		}
\startdata
Standard Experiment & Deep source & Uniform       & $10^{1.5+0.5i}$ s    ~~~$i=1,...,11$&\\
Quench Experiment     & Deep source & Uniform      & Increase towards top    &\\
Photochemical Experiment   & Top source & Non-uniform  & $10^{1.5+0.5i}$ s   ~~~$i=1,...,11$   & 
\enddata
\end{deluxetable*}

For simplicity, we consider a planet with zero obliquity in a circular orbit around its host star. We adopt the radius, mass and rotation period of the canonical hot Jupiter HD209458b. Because we mainly focus on the tracer transport by atmospheric dynamics instead of the general circulation itself, our model setup of the dynamical forcing and damping is essentially the same as that in \citet{liu-showman-2013}. We adopt the Newtonian cooling approximation as a simple radiative heating and cooling scheme, which, when the radiative equilibrium temperature structure and timescale are appropriately chosen, has been shown to produce wind patterns on hot Jupiters (\citealt{liu-showman-2013}) similar as that from GCMs with a realistic radiative transfer scheme (e.g., \citealt{showman-etal-2009}). The heating term in Eq. (8d) is formulated as:
\begin{eqnarray}
\frac{q}{c_p}=\frac{T_{eq}-T(t)}{\tau_{rad}}.
\end{eqnarray}

Motivated by earlier studies using Newtonian cooling (e.g., \citealt{liu-showman-2013}, \citealt{komacek2016atmospheric}, \citealt{zhang2017effects}), we assume that a piecewise vertical profile of the radiative timescale $\tau_{rad}$ that ranges from $10^4$ s at top to $10^6$ s near the bottom (Fig. 1). At pressure levels above 50 Pa, $\tau_{rad}=10^4$ s. At pressure levels below $5\times10^6$ Pa, $\tau_{rad}=10^6$ s. In between, $\tau_{rad}=10^{6}(p/p_0)^{0.4}$ s with $p_0=5\times10^6$ Pa. 
\begin{figure}[t]
  \centering \includegraphics[width=0.48\textwidth]{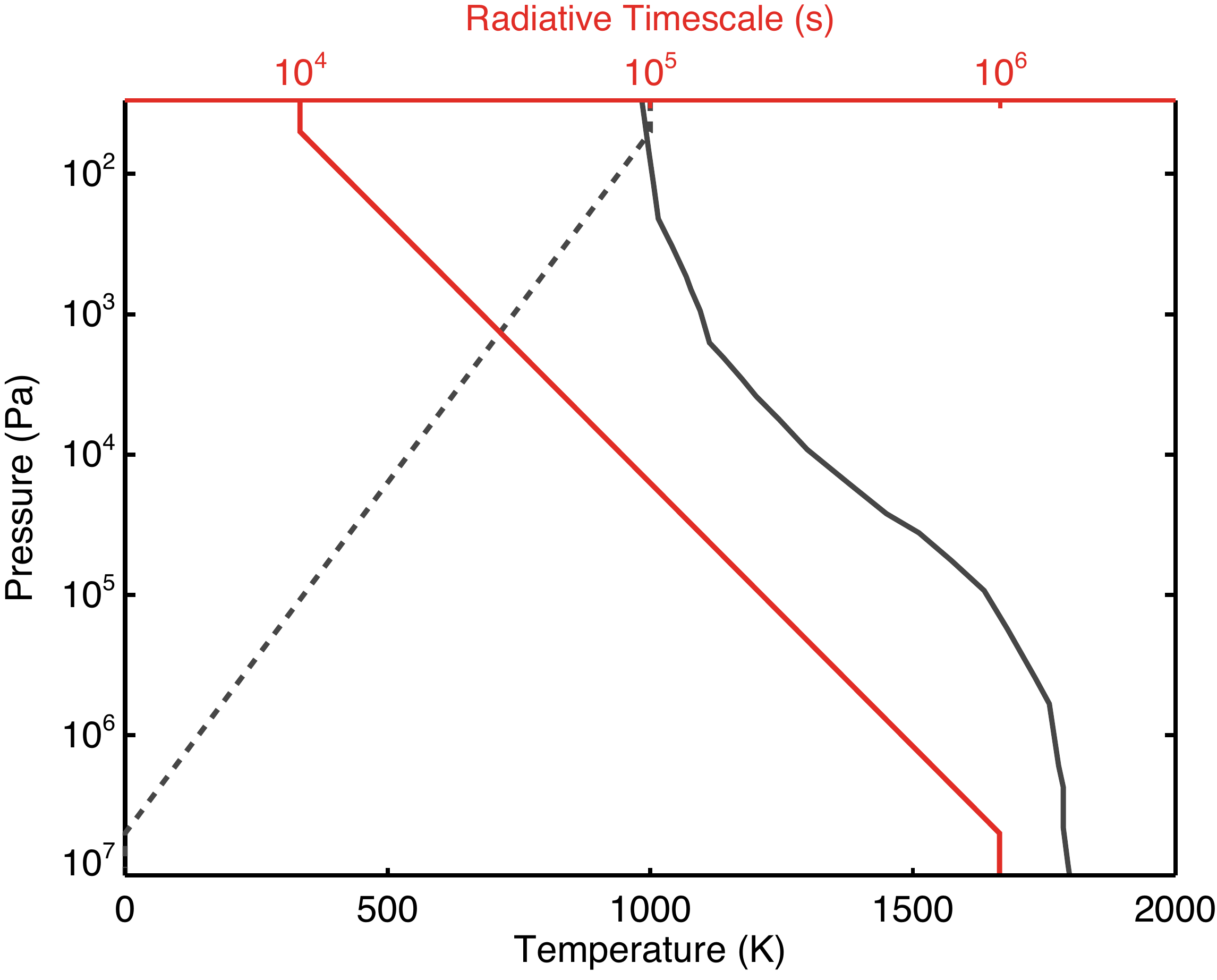} 
  \caption{Vertical profiles of the reference temperature (black), radiative timescale (red) and a typical day-night temperature difference ($\Delta T_{eq}$, dashed) in 3D simulations. $\Delta T_{eq}$ is zero below $5\times10^6$ Pa.} 
\end{figure}

We first define a reference temperature-pressure profile $T_0(p)$, which represents a characteristic planet-wide (global) mean temperature profile that was determined by a 1D radiative-convective calculation in \citet{iro-etal-2005} (Fig. 1). Given this profile, we then analytically construct a nightside equilibrium temperature $T_{n} (p)=T_{0} (p) - \Delta T_{eq}(p)/2$. The equilibrium temperature difference $\Delta T_{eq}$ is chosen as 1000 K at the top of the atmosphere and linearly decreases towards zero at the bottom in the log-pressure coordinate (Fig. 1). Then, the spatial distribution of the radiative equilibrium temperature $T_{eq}$ across the globe is given by:
\begin{gather}
\label{eq:foo}
T_{eq} (\lambda, \phi, p)=
 \begin{cases}
T_{n} (p) + \Delta T_{eq}(p)\cos\lambda\cos\phi &\mathrm{dayside}\\
T_{n} (p) &\mathrm{nightside}
\end{cases}
\end{gather}
where $\lambda$ is longitude and we assume a homogeneous equilibrium temperature on the nightside. 

To approximately parameterize interactions between the atmospheric flow in our simulated domain and a relatively quiescent interior below, we assume a linear ÒbasalÓ drag scheme in the momentum equation (Eq. 8a), following \citet{liu-showman-2013}. $F_d=\vec{\bf{u}}/\tau_{drag}$. The drag coefficient ($\tau_{drag}^{-1}$) is assumed as 0.1 $\mathrm{day^{-1}}$ at the bottom of the domain ($10^7$ Pa) and decreases linearly with decreasing pressure to zero at $5\times10^5$ Pa, above which the atmosphere is essentially drag-free.

As in Paper I, we adopt a linear chemical scheme which relaxes the tracer distribution towards local chemical equilibrium $\chi_0$. Here we performed a single 3D dynamical simulation with three sets of passive chemical tracers. We have 11 tracers in each set as an experiment. In other words, the tracer equation (8f) is actually 33 distinct equations for all tracers. Each tracer is fully and self-consistently coupled to the exact same dynamical wind pattern, but because they are passive, the tracers do not affect the dynamics, and moreover the tracers are all independent of each other, and do not interact. Therefore, one can conceptually view the dynamics, and any given single tracer, as a coupled chemical-dynamical simulation that shows how that tracer would behave, under the influence of the 3D advection field along with the tracer sources and sinks. 

The design of the three sets of tracers are summarized in Table 1. In the first set (``Standard Experiment"), tracers are advected from the deep atmosphere and the distribution of $\chi_0$ is assumed to be horizontally uniform across the globe. We introduce 11 passive tracers of differing chemical timescale $\tau_c$ (Eq. 9), with the timescale of the $i$th tracer specified as $\tau_c=10^{1.5+0.5i}$ s, where $i$ runs from 1 to 11. The chemical timescale is assumed constant with altitude. The vertical profile of $\chi_0$ is a power function of pressure $\chi_{0}=10^{-4}(p/p_0)^{1.6}$ with $p_0=5\times10^6$ Pa. It starts from $10^{-4}$ at bottom and decreases towards $10^{-12}$ near the top. 

In the second tracer set (``Quench Experiment"), we investigate how the tracers are dynamically quenched as they are transported upward from the deep atmosphere. The tracer distribution under chemical equilibrium $\chi_0$ is the same as the ``standard case", i.e., uniform across the globe and decreases towards upper atmosphere. But in this case the chemical timescales of the tracers are relatively long compared with tracers in the Standard case and also increase linearly with decreasing pressure in log-pressure coordinates. The chemical timescale is fixed as $10^4$ s at $5\times10^6$ Pa and that of the $i$th tracer is set as $\tau_c=10^{5.5+0.5i}$ s at 50 Pa. 

In the third tracer set (``Photochemical Experiment"), we simulate photochemically produced species produced in the upper atmosphere on the dayside. A non-uniform chemical equilibrium abundance is prescribed with a significant day-night contrast and with the dayside chemical-equilibrium abundance maximizing at high altitude:
\begin{gather}
\label{eq:foo}
\chi_0 (\lambda, \phi, p)=
 \begin{cases}
\chi_n + \Delta \chi_{eq}(p)\cos\lambda\cos\phi &\mathrm{dayside}\\
\chi_n &\mathrm{nightside}
\end{cases}
\end{gather}
where $\chi_n=10^{-12}$ is the equilibrium mixing ratio on the nightside, which is taken constant here. The equilibrium tracer mixing ratio $\Delta \chi_{eq}=10^{-12}(p/p_0)^{-1.6}$ where $p_0=5\times10^6$ Pa. $\Delta \chi_{eq}$ at the substellar point is $10^{-4}$ near the top of the atmosphere and decreases towards $10^{-12}$ at the bottom. The tracer chemical timescales are set the same as the Standard case and constant with pressure. 

We simulate the atmospheric circulation and tracer transport using the MITgcm (\citealt{adcroft-etal-2004}) dynamical core that solves the primitive equations using the finite volume method on a cubed sphere grid. MITgcm has been extensively used in studying atmospheric dynamics on planets (e.g., \citealt{adcroft-etal-2004}; \citealt{lian-showman-2010}; \citealt{showman-etal-2009}; \citealt{kataria-etal-2014}) and passive tracer transport on a tidally locked exoplanet (\citealt{parmentier20133d}). The horizontal resolution of our simulations is C64 on a cubed sphere grid, corresponding to 256$\times$128 in longitude and latitude, or $1.4^\circ$ per grid cell. We also tested some cases using the C128 grid, corresponding to a global resolution of approximately 512$\times$256 in longitude and latitude, or $0.7^\circ$ per grid cell. The results are consistent with our lower-resolution runs. We use 80 levels from about $10^7$ Pa to $30$ Pa, evenly spaced in log pressure. The time step is 20 second. We apply a fourth-order Shapiro filter to the time derivatives of horizontal velocities and potential temperature with a damping time-scale of 25 second to ensure the numerical stability. The passive tracer transport is solved using a second-order flux-limiter advection scheme. The simulations were performed for at least 5000 Earth days in model time to ensure the statistical properties discussed here have reached an equilibrated state. All variables were averaged over the last 400 days for analysis. 

\subsection{Results: Standard Experiment}

\begin{figure*}
\includegraphics[width=0.95\textwidth]{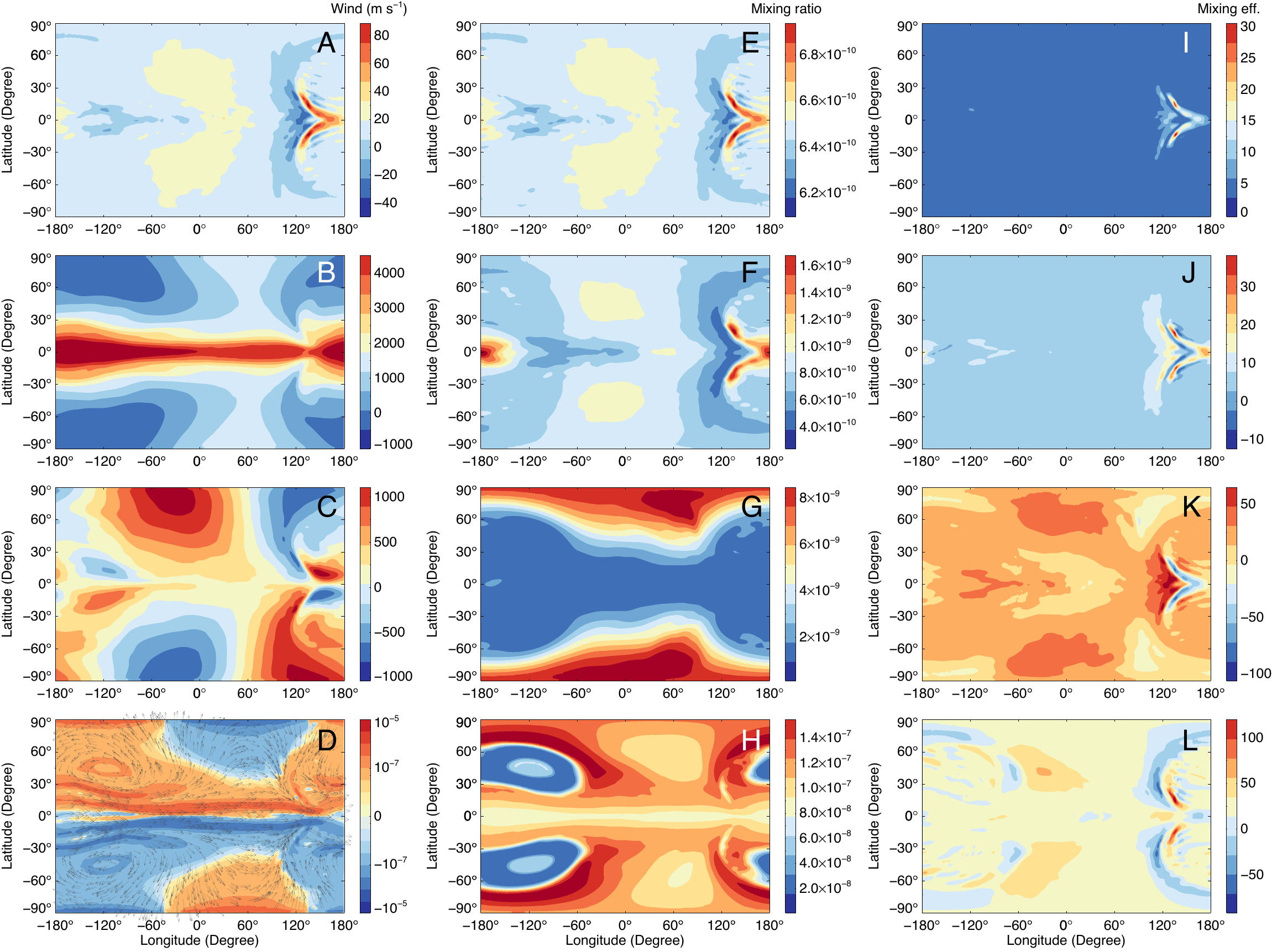} 
 \centering\caption{Longitude-latitude maps of various quantities at 2800 Pa in the Standard Experiment. Left column from top to the bottom (Panel A-D): vertical wind ($w$), east-west wind ($u$), north-south wind ($v$), and relative vorticity ($\zeta$) with wind field vectors. Central column (E-H): volume mixing ratios of tracers with chemical timescale of $10^{2.5}$ s, $10^{4}$ s, $10^{5.5}$ s, $10^{7} $s from top to the bottom, respectively. Right column (I-L): corresponding vertical mixing efficiency of the tracers in the central column. Note that the color bars do not have the same range.} 
\end{figure*}

Under a steady day-night forcing, the 3D wind pattern from our simulations is consistent with previous studies on typical hot Jupiters (e.g., \citealt{showman-etal-2009}). In the east-west direction, an equatorial super-rotating wind reaches 3 to 4 $\mathrm{km~s^{-1}}$ at about 3000 Pa (Fig. 2B). A weak westward wind develops at high latitudes on the nightside. In the north-south direction, a large-scale meridional wind blows from low latitude to the high latitude on the dayside, and the flow returns to low latitudes on the nightside (Fig. 2C). The magnitude of the meridional wind is about several hundred $\mathrm{m~s^{-1}}$. 

At low latitudes, an outstanding feature of the vertical wind pattern is the adjacent strong upwelling and downwelling flows on the nightside between the east terminator and the substellar point, i.e., longitude between 90 to 180 degrees (Fig. 2A). This is consistent with several prior investigations showing such a feature (e.g., \citealt{showman-etal-2009}; \citealt{rauscher-menou-2010}). The upwelling and downwelling flows correspond to the local divergence and convergence of the horizontal flows. A typical upwelling flow forms a chevron-like pattern tilted towards the northwest-southeast direction in the northern hemisphere and towards the southwest-northeast direction in the southern hemisphere. Surrounding the upwelling flows are the downwelling flows with a magnitude up to 20-40 $\mathrm{m~s^{-1}}$. In the region between longitudes -60 and 60 degrees, the dayside is dominated by a large, weaker upwelling flow pattern in both the northern and southern hemispheres. Correspondingly, the meridional wind flows away from equator with an accelerating speed that maximizes at high latitude, leading to a large-scale divergence on the dayside (Fig. 2C). The advection timescale due to the strong equatorial super-rotating wind $a/U$ is on the order of $10^4$ s, and that due to the meridional wind $a/V$ is on the order of $10^5$ s.

\begin{figure}[t]
\includegraphics[width=0.48\textwidth]{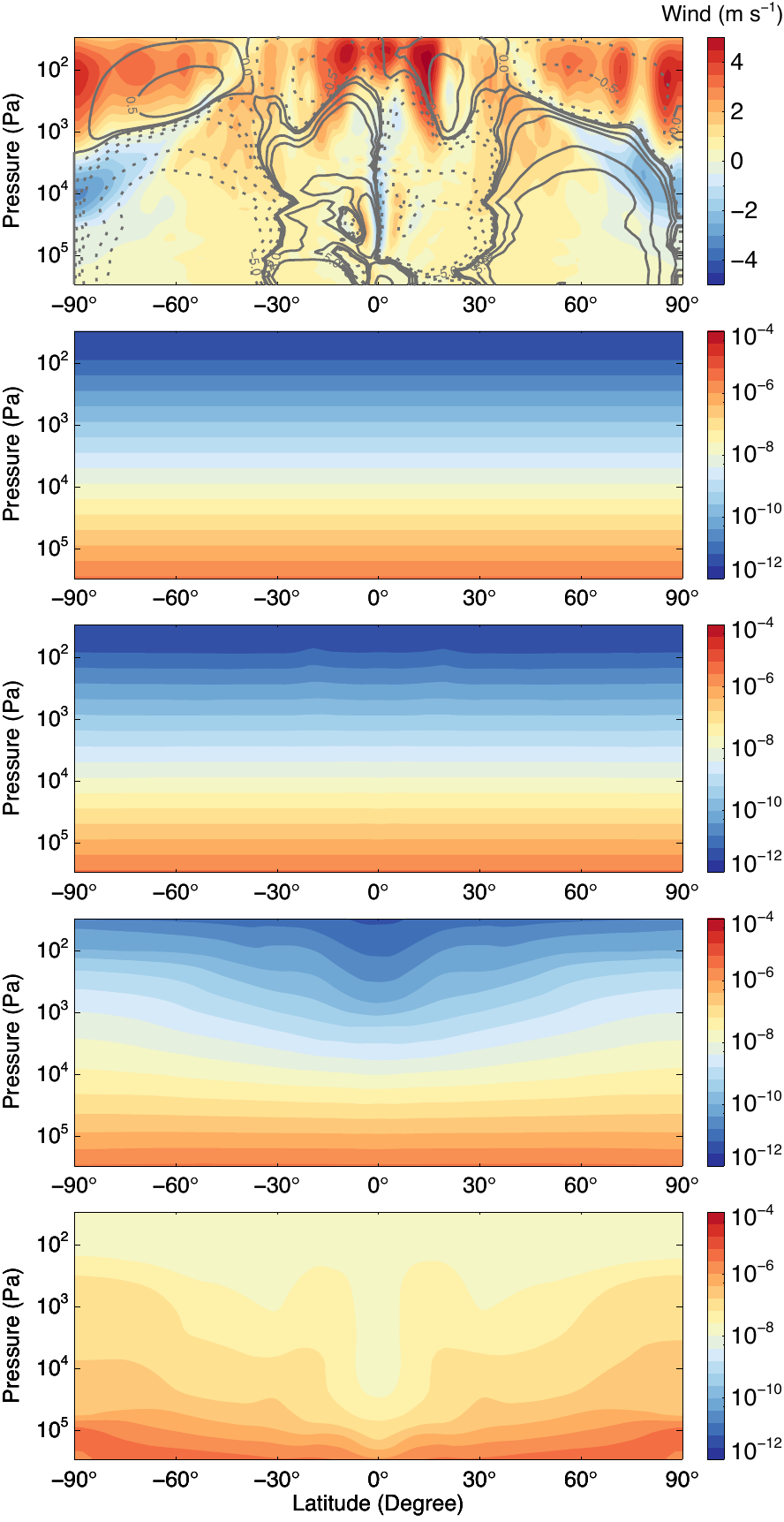} 
 \centering\caption{Latitude-pressure maps of zonally averaged quantities in the Standard Experiment. Top: vertical wind velocity (color) and TEM mass streamfunction (contours, in units of $10^{12}$ Kg $\mathrm{s^{-1}}$, solid/positive is clockwise). Starting from the second row we show volume mixing ratio of tracers with different chemical timescales of $10^2$ s, $10^{3.5}$ s, $10^5$ s, $10^{6.5}$ s from top to bottom, respectively.} 
\end{figure}

Very short-lived species generally exhibit small spatial deviations from their uniform equilibrium distribution. As predicted in Paper I, the spatial distribution of a short-lived tracer (Fig. 2E) follows the vertical velocity pattern (Fig. 2A). Two large-scale positive anomalies are located in the northern and southern hemispheres on the dayside, respectively. Adjacent positive and negative anomalies are found near the equatorial region on the nightside between the east terminator and the substellar point, corresponding to the upwelling and downwelling flows, respectively. As the chemical lifetime of the species increases, the correlation between the tracer and vertical velocity pattern becomes weaker. For species with a chemical timescale longer than advection timescales ($10^4-10^5$ s), the tracer tends to accumulate in the middle and high latitudes (Fig. 2G). Its longitudinal variation is generally homogenized and is much smaller than the latitudinal variation. For even longer-lived species ($\tau_c>10^7$ s), the horizontal tracer distribution (Fig. 2H) exhibits some similarities with that of the relative vorticity $\zeta=\nabla\times\vec{\bf{u}}$ where $\vec{\bf{u}}$ is the horizontal wind velocity vector (Fig. 2D), suggesting the spatial pattern of inert tracer is significantly controlled by the horizontal flow pattern. 

To quantify the local contribution of vertical tracer transport, we calculate the ``mixing efficiency'' $\eta$ for each tracer (\citealt{parmentier20133d}):
\begin{eqnarray}
\eta=\frac{w\chi-\overline{w\chi}}{\overline{w\chi}}.
\end{eqnarray} 

Note that $\overline{w\chi}=\overline{w\chi^\prime}$ is the global-mean vertical tracer flux. $\eta$ represents the deviation of the local tracer flux from the global mean and normalized by the mean value. For a short-lived species, due to the strong correlation between the tracer and vertical velocity, mixing efficiency is positive everywhere on an isobar (Fig. 2I). $\eta$ strongly maximizes at the equator on the nightside between the east terminator and the anti-stellar point, producing localized plumes or ``tracer chimneys'' (\citealt{parmentier20133d}). That the vertical tracer transport is significantly controlled by the local upwelling and downwelling plumes on tidally locked planets is consistent with \citet{parmentier20133d}. However, as the tracer chemical lifetime increases, vertical mixing becomes a bit complicated. First, negative $\eta$ emerges due to imperfect correlation (or some local anti-correlation) between the tracer distribution and vertical velocity (Fig. 2J). Second, the localized ``tracer chimneys'' where efficient transport takes place across isobars in the short-lived tracer regime become less pronounced as the tracer becomes longer-lived (Fig. 2K). Interestingly, the strong upwelling plume near the east terminator, which is mostly responsible for the upward mixing for short-lived species, turns to cancel the upward mixing in the long-lived species case as the horizontal tracer distribution changes with the tracer chemical lifetime (Fig. 2L).

The zonally averaged vertical velocity and tracer abundance are shown in Fig. 3. The zonal-mean circulation exhibits a net downwelling at the equator with two strong upwelling regions at low latitudes off the equator. The polar regions are mostly downwelling. Correspondingly, the Transformed Eulerian Mean (TEM) streamfunction exhibits two ``anti-Hadley cells" in the equatorial region\footnote{This result seems consistent with a recent Super-Earth study by \citet{charnay20153d}. But here we emphasize that whether this ``anti-Hadley" circulation occurs depends on the parameters in the setup. Some hot Jupiter simulations do not show this feature (personal communications with V. Parmentier)}. Two more circulation cells exist with ascending branches at middle latitudes and descending branches at high latitudes. As a result, the zonal-mean distribution of long-lived tracers exhibits a local minimum near the equator in our simulations (Fig. 3). 

Similar to the 2D model results, the global-mean tracer mixing ratio profile is close to the chemical equilibrium if the chemical timescale is short and becomes more vertically homogenized as the tracer lifetime increases (Fig. 4).  A detailed discussion of the departure of the tracer vertical profile from its chemical equilibrium profile will be presented later in the ``Quench Experiment". 

Based on the global-mean tracer distribution from simulations, we numerically derived effective eddy diffusivities $K_{zz}$ from the flux-gradient relationship (e.g. Plumb and Mahlman 1987):
\begin{eqnarray}
\overline{w\chi^\prime}\approx -K_{zz}\frac{\partial\overline{\chi}}{\partial z}.
\end{eqnarray}
$K_{zz}$ increases with decreasing pressure as the vertical velocity also increases. $K_{zz}$ increases with chemical timescale, spanning over 3-4 orders of magnitude. However, for long-lived species, the dependence of $K_{zz}$ on chemical lifetime is complicated. At some pressure (e.g., 2000 Pa), as the tracer lifetime increases, $K_{zz}$ actually decreases in this regime. Again, this behavior can be understood owing to the fact that the material surfaces of the tracer have been significantly distorted in the case of long-lived species. The correlation between the vertical velocity and tracer is not good (Fig. 2K and 2L) and eventually our theory in Section 2 based on relatively short-lived species fails in this regime.

\begin{figure}[t]
\includegraphics[width=0.48\textwidth]{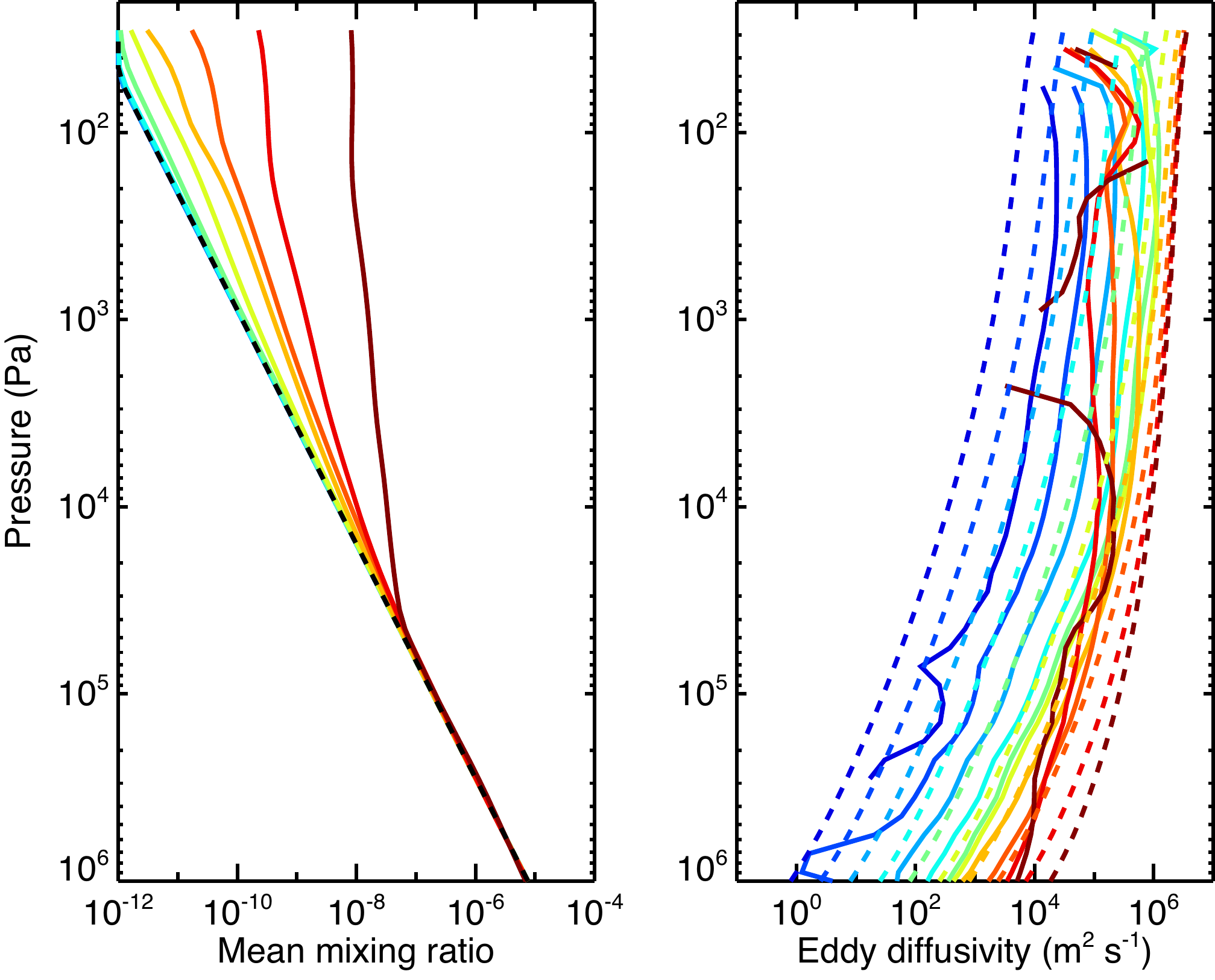} 
 \centering\caption{Left: Vertical profiles of the global-mean volume mixing ratio in the Standard Experiment. Right: vertical profiles of numerically derived effective eddy diffusivity $K_{zz}$ (right) from the flux-gradient relationship (Eq. 8) based on simulations (solid). The dashed lines are analytical predictions using Eq. (1) and the analytical $\hat{w}$ in Eq. (7). Different colors from cold (blue) to warm (red) represent tracers with different chemical timescales from short to long, ranging from $10^2$ s to $10^{6.5}$ s.} 
\end{figure}

\begin{figure}[t]
\includegraphics[width=0.48\textwidth]{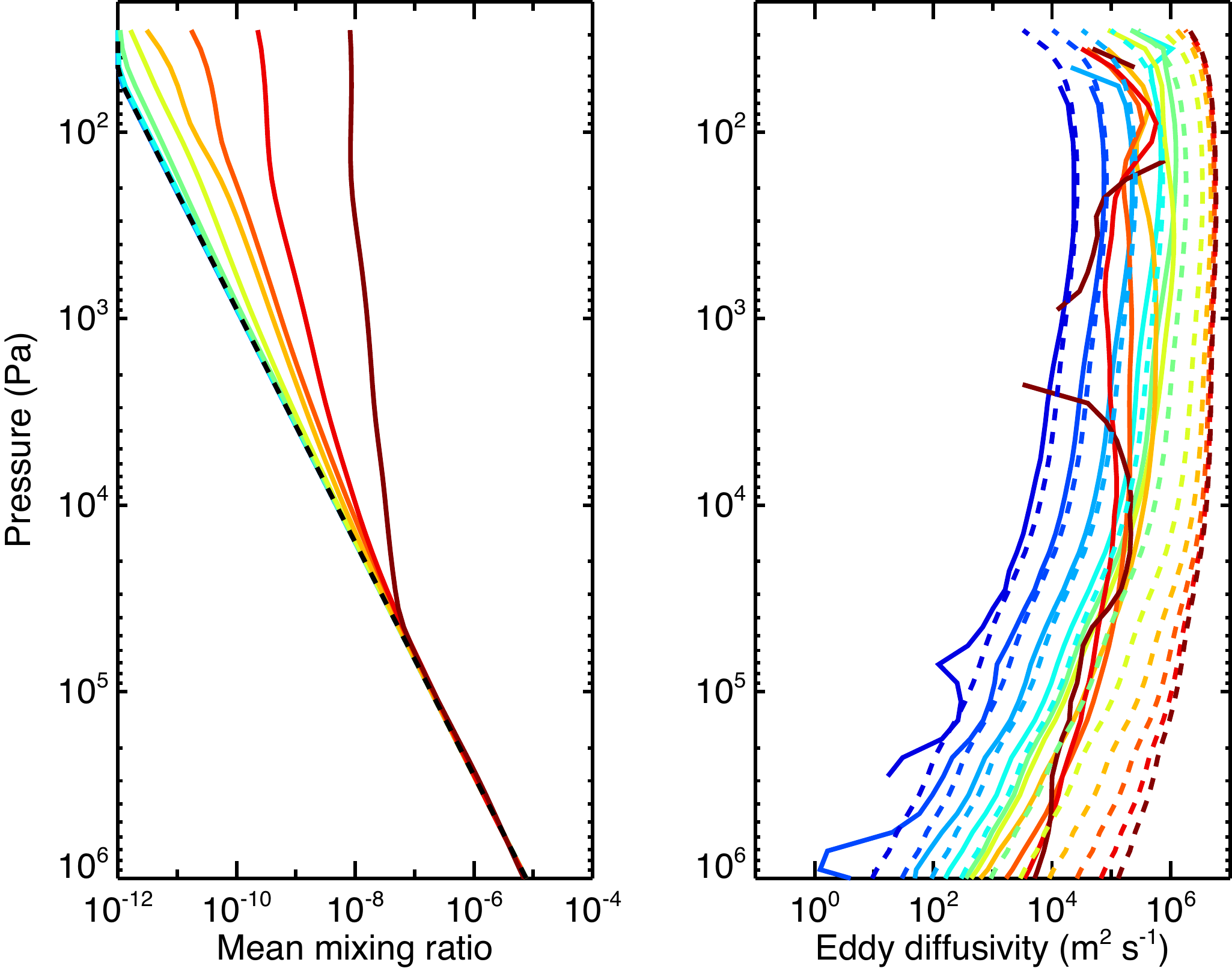} 
 \centering\caption{Same as Fig. 4 but the predicted $K_{zz}$ (dashed) is calculated using Eq. (1) with the vertical velocity directly calculated form the numerical simulations.} 
\end{figure}

We also analytically predict $K_{zz}$ based on the theory in Section 2. As this is a uniform $\chi_0$ case in regime I, the non-diffusive term vanishes. Inserting the expression of $\hat{w}$ (Eq. 7) into Eq. (1), we can estimate the profiles of $K_{zz}$. We assume the vertical characteristic length scale $L_v$ as the pressure scale height $H$ in Eq. (1), which is equivalently to assuming typical horizontal length scale $L_h$ as the planetary radius $a$ according to the continuity equation $U/a\sim \hat{w}/H$. 

The theoretically predicted $K_{zz}$ are shown in Fig. 4. Similar to the numerically derived $K_{zz}$, the predicted $K_{zz}$ increases with the chemical lifetime and with decreasing pressure. For most tracers, the analytical $K_{zz}$ matches well with the $K_{zz}$ derived from the simulations. This implies that our theory in Section 2 applies generally well to the regime of 3D atmospheres on close-in tidally locked exoplanets with a uniform source that has greatest chemical equilibrium abundance at depth.

Rigorously, our theory does not perform well in either very-short-lived or very-long-lived tracer regimes (Fig. 4). The analytical theory underestimates the numerical $K_{zz}$ by about a factor of 10 for short-lived tracers. This discrepancy is because both the horizontal (Eq. 4) and vertical (Eq. 7) velocities estimated from our analytical theory are less by about factor of 2-3 than the numerical values in a drag-free hydrogen atmosphere (\citealt{komacek2016atmospheric}, \citealt{zhang2017effects}). When the tracer lifetime is very small, i.e., $\tau_c\rightarrow 0$, $K_{zz}\approx \hat{w}^2\tau_c$ in Eq. (1), i.e., $K_{zz}$ scales with the vertical velocity squared. Therefore $K_{zz}$ predicted by our theory is about 10 times smaller than that from 3D models for the short-lived tracers. When the drag is applied in the model, the analytical $K_{zz}$ matches better with the numerical simulations (personal communication with T. Komacek). If we directly use $\hat{w}$ from our numerical simulations, rather than using the analytically predicted vertical velocities, the agreement is improved in the short-lived tracer regime (Fig. 5). 

However, as shown in Fig. 4 and 5, our $K_{zz}$ theory generally overestimates the $K_{zz}$ if the tracer lifetime is long. This is Regime III where the tracer contours are significantly distorted. For 2D simulations in Paper I, we also found that the $K_{zz}$ theory does not behave well for long-lived tracers. But in the 2D cases the analytical $K_{zz}$ generally underestimates the eddy mixing caused by the resolved circulation in the 2D model.  Again, the reason could be the adopted tracer transport length scale here---we have assumed $L_v\sim H$ and $L_h\sim a$. But as shown in the mixing efficiency maps (Fig. 2), strong local tracer transport could dominate the entire globe, implying a smaller $L_h$. As also mentioned in Section 2, in the long-lived tracer regime (regime III), $L_v$ might be different from the atmospheric scale height $H$. Previous studies (e.g. \citealt{cooper-showman-2006}) have pointed out that $L_v$ for CO transport on hot Jupiters might be a factor of 2-3 smaller than $H$. If we adopt a smaller dynamical length scale, the resultant $K_{zz}$ would be smaller and match the 3D numerical results better in the long-lived tracer regime. 

\subsection{Results: Quench Experiment}

For tracers that are transported from deep atmospheric sources, their mixing ratios can be vertically homogenized by the atmospheric dynamics and ``quenched" at a value at a deeper pressure level (Fig. 6). This usually occurs when the chemical timescale of the tracer changes significantly with pressure. For example, high temperature in the deeper atmosphere maintains fast thermochemical reactions of a tracer and results in a short chemical timescale. Chemistry dominates over the dynamics and the tracer is expected to be in local thermochemical equilibrium. As the tracer is advected upward, chemical reactions slow down as the temperature decreases in the upper atmosphere, and the chemical timescale increases quickly. At a certain pressure level, the vertical profile of the chemical tracer starts to depart from its chemical equilibrium profile. Above that level, the tracer will be more controlled by the atmospheric dynamics. If the tracer chemical timescale increases so rapidly with decreasing pressure that the chemistry can be quickly neglected above the departure level (e.g., carbon monoxide on Jupiter, \citealt{prinn1977carbon}), the tracer mixing ratio in the upper atmosphere is approximately constant and is basically set by the value at the departure level. This is called quenching. 

\citet{bordwell2018convective} defined two kinds of ``quench points". The first one is the ``departure" quench point, which stands for the pressure level where the tracer vertical profile starts to deviate from the chemical equilibrium tracer profile. The second one is the ``observed" quench point, referring to a pressure level at which the equilibrium tracer mixing ratio is equal to the tracer mixing ratio in the upper atmosphere. There is another ``quench points" that was not mentioned in \citet{bordwell2018convective}, the ``reflection point", where the actual tracer vertical profile becomes vertically constant\footnote {One should not confuse the ``reflection point" with the ``observed" quench point. The reflection point refers to the pressure level on the actual mixing ratio profile where the tracer vertical profile becomes vertically constant, while the ``observed" quench point refers to the pressure level on the equilibrium mixing ratio profile where the equilibrium mixing ratio is the same as the observed value in the upper atmosphere.}. The three quench points are the same if the tracer mixing ratio profile becomes vertically constant quickly as it deviates from the equilibrium profile (e.g., red line in Fig. 6). On the other hand, if the chemical timescale does not increase very quickly and the chemistry above the departure point cannot be neglected, the tracer mixing ratio will keep decreasing and may not reach a constant vertical profile until in the very top atmosphere (e.g. blue line in Fig. 6). In this case, the three quench points are not the same, and usually the ``observed" quench point is located between the reflection point and the departure point (Fig. 6). In this study we mainly focus on the departure quench point. 

\begin{figure}[t]
\includegraphics[width=0.48\textwidth]{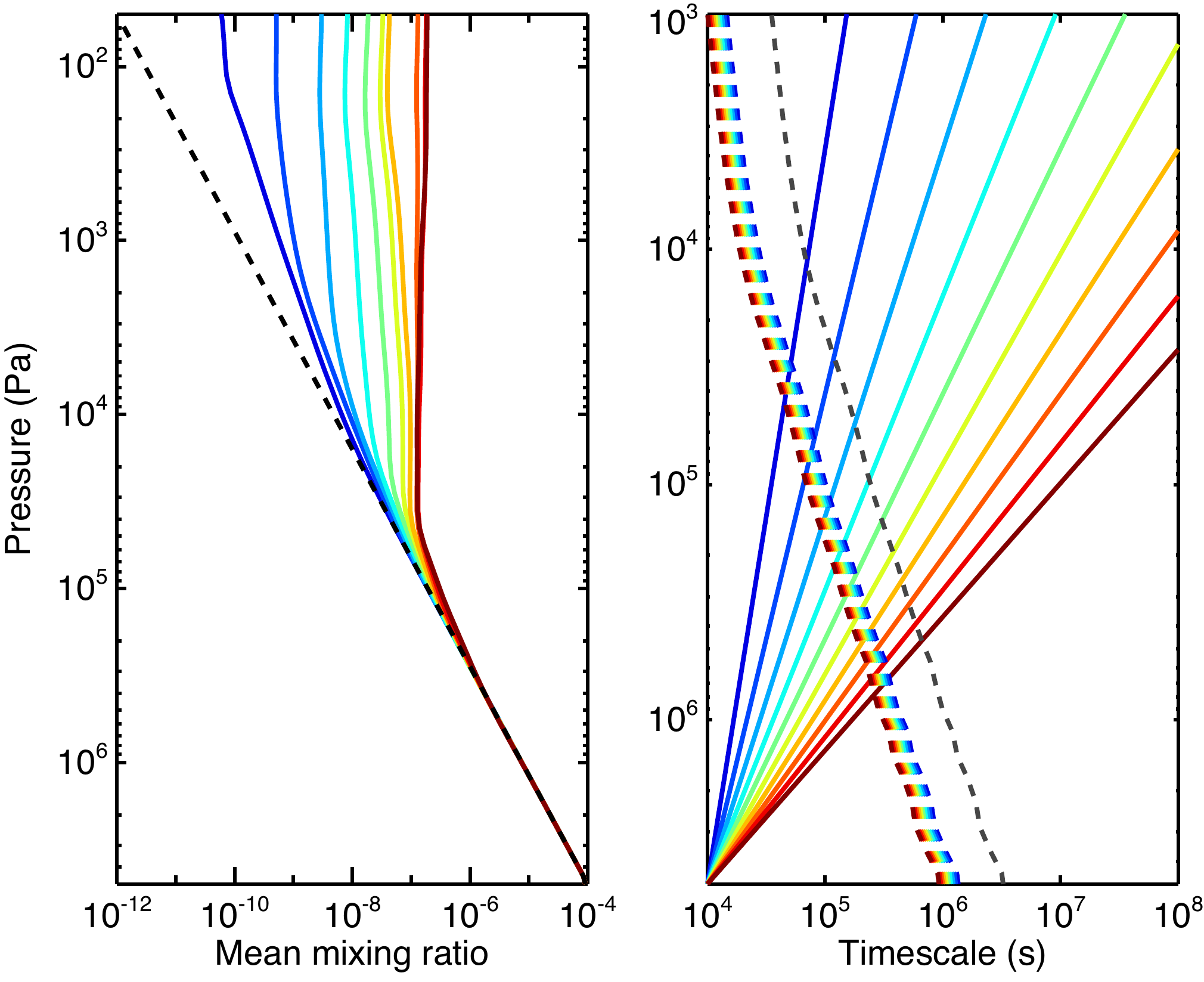} 
 \centering\caption{Left: vertical profiles of the global-mean volume mixing ratio from numerical simulations (solid) and the equilibrium mixing ratio (dashed) in the Quench Experiment. Right: vertical profiles of the chemical lifetimes (solid) and vertical transport timescales based on pressure scale height (black dashed) and equilibrium chemical scale height for individual tracers (color dashed). Different colors from cold (blue) to warm (red) represent tracers with different chemical timescales from short to long.} 
\end{figure}

\begin{figure*}
\includegraphics[width=0.95\textwidth]{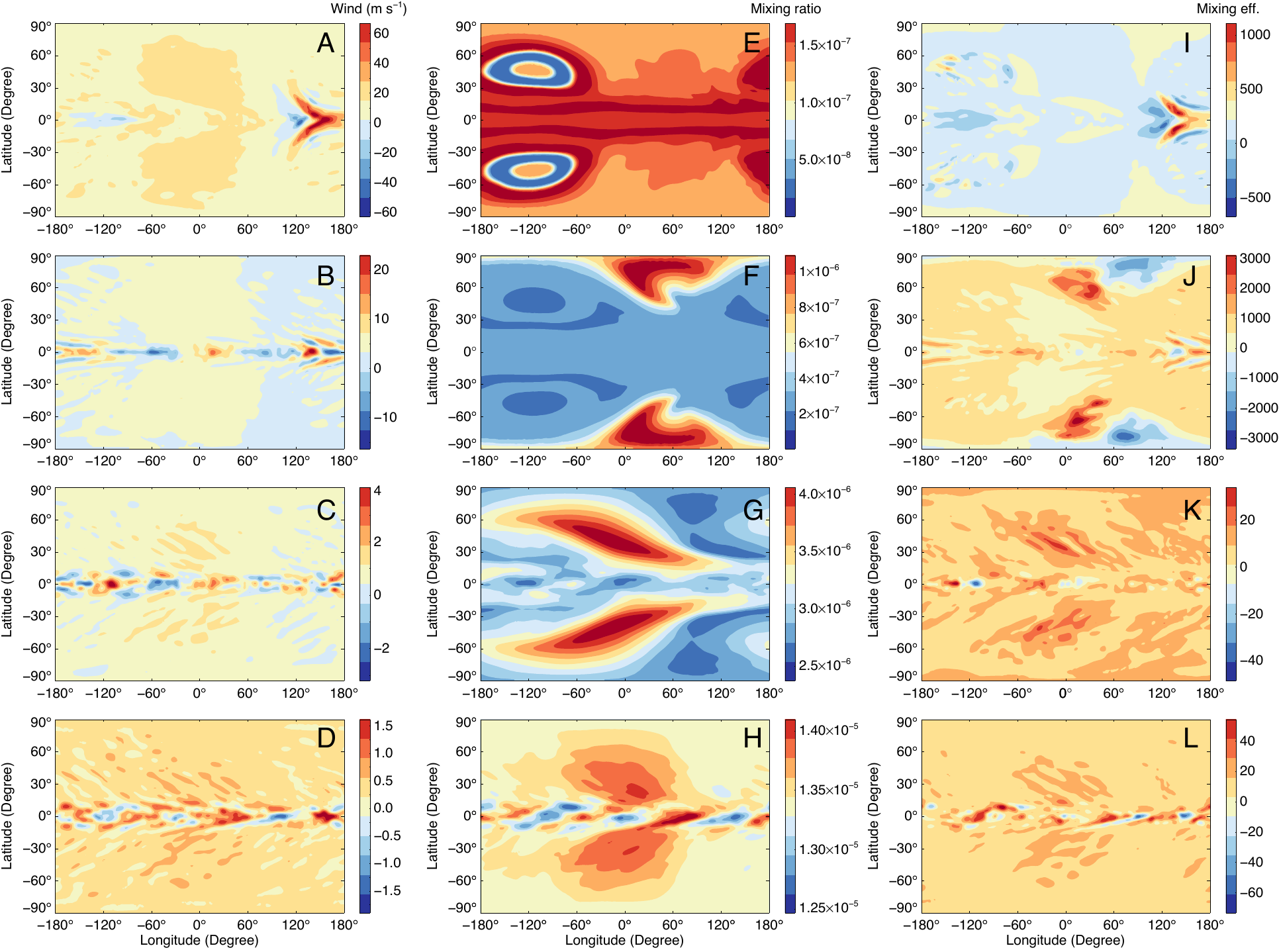} 
 \centering\caption{Longitude-latitude maps of vertical wind (left), tracer mixing ratios (central) and vertical mixing efficiency (right) from 3D Quench simulation of a tracer with chemical timescale $10^{9.5}$ s at $50$ Pa (orange line in Fig. 6). The pressure levels from top to bottom: $9\times10^3$ Pa (upper atmosphere), $5.7\times10^4$ Pa (reflection point), $3.6\times10^5$ Pa (departure point), $1.4\times10^6$ Pa (deep atmosphere).} 
\end{figure*}

Dynamical quenching of the disequilibrium chemical tracers is usually investigated in the convective atmospheres (e.g., \citealt{prinn1976chemistry}, \citealt{prinn1977carbon}, \citealt{smith1998estimation}, \citealt{visscher-moses-2011}, \citealt{bordwell2018convective}). However, in the radiative zone of the atmospheres on tidally locked planets with vigorous vertical transport, dynamical quenching of long-lived species such as $\mathrm{H_2O}$, $\mathrm{CH_4}$ and $\mathrm{CO}$ is also important (e.g., \citealt{cooper-showman-2006}, \citealt{moses-etal-2011}, \citealt{line2011thermochemical}). In our Quench Experiment, the tracer chemical timescales increase with decreasing pressure and span many orders of magnitude (Fig. 6), as expected to occur for a variety of real chemical species, including CH4/CO interconversion (e.g., \citealt{cooper-showman-2006}). The chemical timescale of the longest-lived tracer has a very steep vertical gradient, ranging from $10^4$ s at $5\times10^6$ Pa to $10^{11}$ s at $50$ Pa. The departure point of this tracer from the equilibrium chemical profile is at ~$4\times10^5$ Pa and the ``observed" quench point is roughly at $10^5$ Pa (Fig. 6). The chemical timescale of the shortest-lived tracer ranges from $10^4$ s at $5\times10^6$ Pa to $10^6$ s at $50$ Pa. The departure point is at about $6\times10^4$ Pa. But the tracer mixing ratio decreases with pressure towards the top of the atmosphere because the local chemistry plays a non-negligible role in all pressure range. There is no clear ``observed" quench point for this tracer case.

To understand the details of the dynamical quenching in the atmosphere of a 3D tidally locked planet, we analyzed a typical long-lived tracer with a quenched vertical profile (orange line in Fig. 6). The chemical timescale of this tracer ranges from $10^4$ s at $5\times10^6$ Pa to $10^{9.5}$ s at $50$ Pa. The departure point of this tracer is at about $3.6\times10^5$ Pa and the reflection point is at about $5.7\times10^4$ Pa (Fig. 6). We plotted spatial maps of the vertical wind and tracer mixing ratio at several typical pressure levels (Fig. 7), including an upper atmosphere level ($9\times10^3$ Pa) above the reflection point and a deep atmosphere level ($1.4\times10^6$ Pa) below the departure point. 

There are many similarities in the spatial distributions between Fig. 7 for this typical tracer in the Quench Experiment and Fig. 2 for four different tracers with chemical timescales constant with pressure in the Standard Experiment. This implies that the spatial distributions of the tracer in the Quench Experiment can be approximately understood using its local chemical timescale. Because the tracer chemical timescale increases from bottom to the top, while the vertical transport timescale decreases, the ratio of the chemical timescale and dynamical timescale in the atmosphere $\tau_c/\tau_d$ increases with decreasing pressure. As a result, the tracer in the Quench Experiment at the deep atmosphere level behaves as a short-lived tracer and that at the upper atmosphere level behaves as a long-lived tracer in the Standard Experiment. 

At deep atmospheric levels, the tracer spatial distribution is similar to the vertical wind pattern, especially in the equatorial region (Fig. 7D and 7H). At the departure level, the dynamical timescale is roughly equal to the chemical timescale, and the tracer is more horizontally homogenized (Fig. 7G). Below the departure level, the mixing efficiency maps are generally positive in the off-equatorial region (Fig. 7L), indicating a decent correlation between the tracer distribution and the vertical wind, although at the equator, negative mixing efficiency appears at several locations. At the reflection point where the tracer mixing ratio stops decreasing with decreasing pressure, large positive and negative anomalies show up in the mixing efficiency map (Fig. 7K), implying a large cancellation between the tracer upwelling and downwelling. The tracer is further more spatially homogeneous because atmospheric dynamics becomes more important. Above the reflection point, the tracer distribution (Fig. 7F) looks similar to the typical long-lived tracer in the Standard Experiment (Fig. 2H). Although the $\tau_c/\tau_d$ is large in the upper atmosphere, the tracer is not completely homogenized (Fig. 7E). Instead, its spatial pattern is significantly affected by the horizontal wind dynamics. The local anomalies located at the mid-latitudes near the west terminator (Fig. 7E) are a response to the large local wind gyres seen in the vorticity map (Fig. 2D). The upwelling and downwelling shown in the mixing efficiency map almost cancel out each other (Fig. 7I), resulting in nearly zero net vertical transport flux. From the tracer continuity equation (Eq. 8f), if the local chemical source and sink were neglected ($S\approx 0$), the vertical tracer flux divergence should vanish as well. In the diffusive framework, this implies that the vertical gradient of the tracer mixing ratio is nearly zero, and the tracer profile is approximately constant with pressure (Fig. 6). 

To visualize the vertical transport of this tracer, we also diagnosed the zonal-mean map and meridional-mean map of vertical wind and tracer mixing ratio (Fig. 8). In the zonal-mean maps, we simply averaged the quantities over longitude at each latitude to produce the latitude-pressure distributions. In the meridional-mean maps, at each longitude point, we averaged the quantities over latitude in an area-weighted fashion. The vertical tracer quenching occurs at a deeper atmospheric level in the equatorial region but the quenched tracer mixing ratio is higher than the other latitudes. There appears to be another quench point at around $10^4$ Pa. The equatorial minimum feature in the latitude-pressure distribution is a result of the net downwelling of the lower-mixing-ratio tracers from the upper atmosphere due to the ``anti-Hadley cells" in the equatorial region, as shown in the TEM streamfunction (Fig. 8). We have discussed this behavior in the Standard Experiment (Fig. 3). 

\begin{figure}[t]
\includegraphics[width=0.48\textwidth]{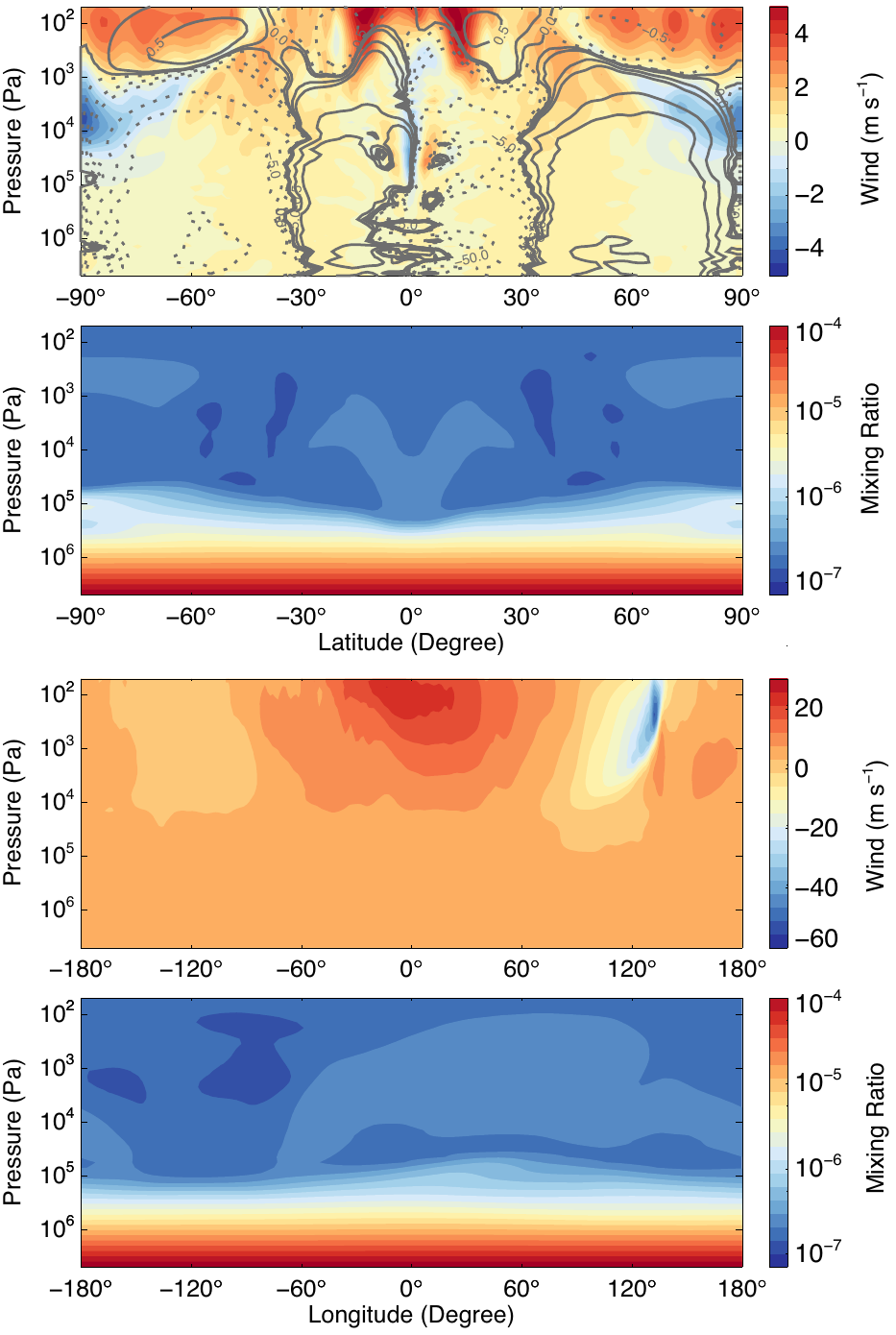} 
 \centering\caption{2D maps from a typical Quench Experiment with chemical tracer timescale $10^{9.5}$ s at $50$ Pa (orange line in Fig. 6). Top: zonally averaged vertical wind velocity (color) and TEM mass streamfunction (contours, in units of $10^{12}$ Kg $\mathrm{s^{-1}}$, solid/positive is clockwise). Second: zonally averaged tracer mixing ratio. Third: area-weighted latitudinal-averaged vertical wind velocity. Bottom: area-weighted latitudinal-averaged tracer mixing ratio.} 
\end{figure}

\begin{figure}[t]
\includegraphics[width=0.48\textwidth]{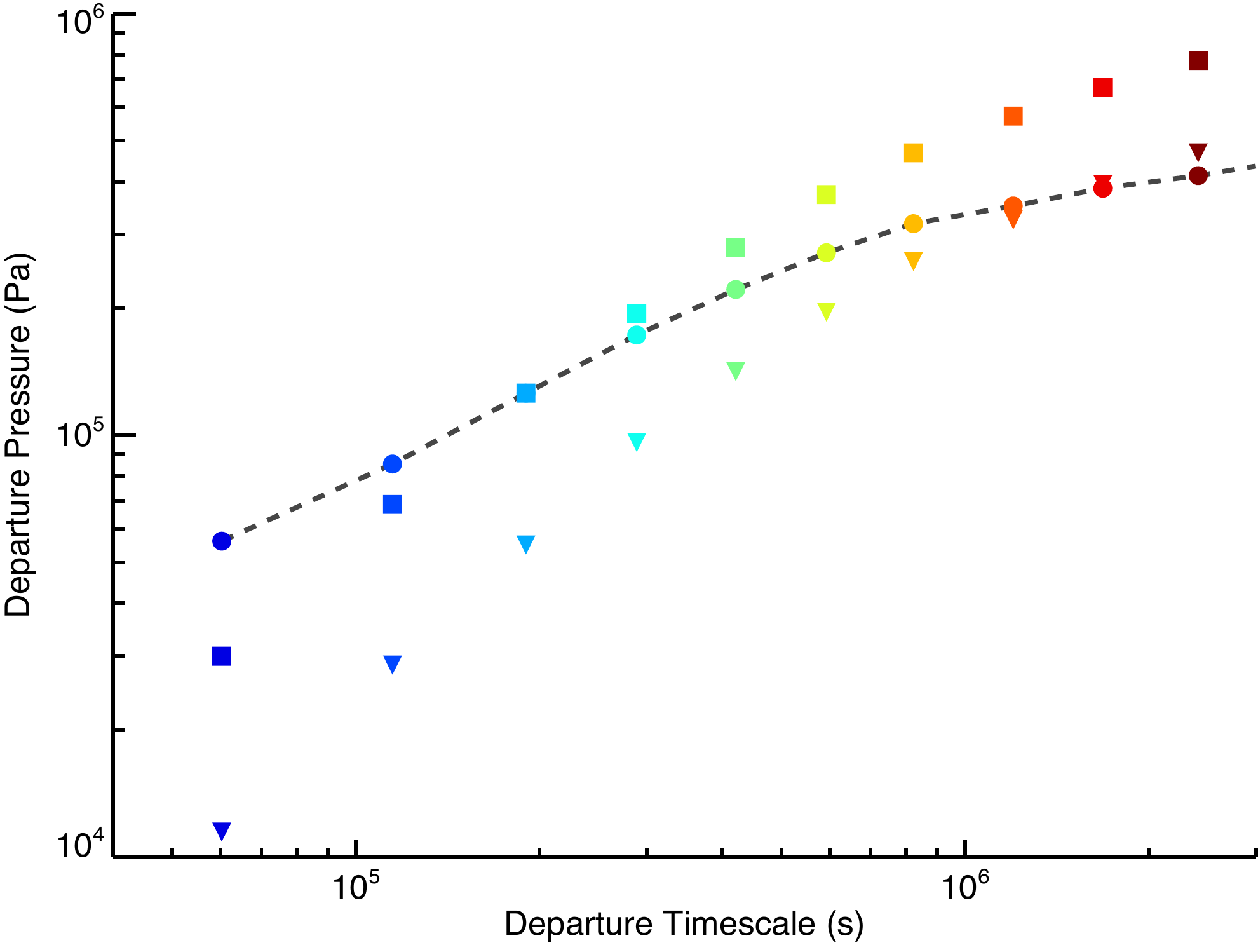} 
 \centering\caption{Departure points as a function of departure temperature in the Quench Experiment (dots connected by the dashed line). Theoretical predictions using our quench theory (Eq. 14) with $L_v\sim H$ (triangles) and $L_v\sim H_{ceq}$ (squares). The color scheme is the same as in Fig. 6.} 
\end{figure}

In the longitudinal direction, the vertical quenching occurs at larger pressure level in the west hemisphere than in the east hemisphere (see the bottom two meridional-mean map in Fig. 8). Furthermore, the quenched tracer mixing ratio in the east hemisphere is actually higher than that in the west hemisphere. The reason is that, when the tracer is transported upward near the substellar longitude from the deep atmosphere, the equatorial superrotating wind advects the tracer to the east hemisphere and enhances the local tracer mixing ratio.

Despite that the quenching level has some spatial variation, here we estimate the global-mean quenching level. In this study we mainly focus on the departure quench point as a more generic metric. We determined the departure points from the tracer profiles in the Quench simulations (Fig. 9). The numerical departure points are measured as the pressure level where the simulated tracer mixing ratio deviates from (larger than) the equilibrium mixing ratio by a small fraction (Fig. 6). In this study we use 1\%. As the tracer mixing ratio increases very quickly above the departure point, this estimated departure point is not very sensitive to the choice of the small fraction (e.g., 10\%). 

Crudely speaking, quenching (or departure) occurs where the chemical timescale is equal to the vertically mixing timescale (e.g., \citealt{prinn1977carbon}, \citealt{cooper-showman-2006},  \citealt{visscher-moses-2011}). In the global-mean vertical eddy diffusion framework, the vertical mixing timescale is the effective diffusive timescale. In the 3D atmosphere with uniform $\chi_0$, Eq. (1) is a good approximation of the eddy diffusivity. As our theory of $K_{zz}$ does not make any assumption whether $\tau_c$ should be constant with pressure or not, Eq. (1) is also applicable to the Quench Experiment in which tracer lifetime varies with pressure. Equating the chemical timescale and effective diffusion timescale at the quench point, we can achieve:
\begin{equation}
\tau_c\sim\frac{L_v^2}{K_{zz}}\approx\frac{L_v}{\hat{w}}(1+\frac{L_v}{\hat{w}\tau_{c}})
\end{equation}
where $\tau_c\sim L_v/\hat{w}$ is the traditional quenching level estimate used in the convective atmosphere (e.g.,\citealt{prinn1977carbon}) and on hot Jupiters (e.g., \citealt{cooper-showman-2006}). Our quenching theory has an additional term that represents the effect of the tracer chemistry on the global-mean vertical tracer transport efficiency. This effect has not been addressed before because previous quenching theories did not include large-scale circulation to estimate the global-mean tracer transport. Previous studies (e.g., \citealt{smith1998estimation}, \citealt{cooper-showman-2006}, \citealt{bordwell2018convective}) showed that the tracer chemistry could affect the vertical characteristic length scale $L_v$, leading to a different departure point estimate. Following \citet{prinn1977carbon}, \citet{bordwell2018convective} proposed a length scale as the scale height of the tracer destruction rate under chemical equilibrium. But in our theory this length scale only depends on the vertical gradient of the chemical timescale, not the chemical timescale itself.

One can actually solve for $\tau_c$ in Eq. (14):
\begin{equation}
\tau_c\sim\frac{1+\sqrt{5}}{2}\frac{L_v}{\hat{w}}.
\end{equation}
It implies that the traditional vertical length scale should be scaled by an additional factor, the golden ratio, to take the tracer chemistry into account. This will increase $L_v$ by a factor of about 1.6. Compared with the conventional estimate using $\tau_c\sim L_v/\hat{w}$, the quench point from our theory is shifted to the pressure level with a larger $\tau_c$. If the tracer chemical timescale increases with decreasing pressure, our predicted quench point is located at higher atmosphere than the conventional estimate. Physically speaking, our global-mean eddy diffusivity theory states that a tracer with a shorter chemical timescale tends to have a smaller $K_{zz}$ than the conventional theory which assumes $K_{zz}\sim\hat{w}L_v$ (Section 2). As a result, using the chemical-dependent eddy diffusivity implies less vertical mixing than the conventional theory, and causes the tracer to be quenched in a higher atmosphere. 

Let us first assume the horizontal dynamical timescale $\tau_d$ is $a/U$ and thus $L_v$ is the pressure scale height $H$, the right hand side in Eq. (15), which we call the ``effective dynamical timescale" in this study, is shown in Fig. 6. We can find out its crossover point with the chemical timescale profile $\tau_c$ to predict the departure point of that species. Using $\hat{w}$ from the 3D simulations, the predicted departure points agree well with the longer-lived tracers in our simulations (Fig. 9). But for shorter-lived species, the predicted quenching pressure is much smaller than the model results. 

It is possible that horizontal dynamical timescale is different from $a/U$ and $L_v$ is not the same as $H$. We tested the situation in which $L_v$ is assumed as the ``chemical scale height", $H_c$, the scale height of the tracer chemical destruction rate that was introduced in the convective atmosphere study (\citealt{bordwell2018convective}). As we use a simple linear chemical scheme in the Quench Experiment, the chemical loss rate can be expressed as $N\bar{\chi}/\tau_c$, where $N$ is the atmospheric number density and $\bar{\chi}$ is the global-mean vertical profile of tracer. The chemical scale height in our study is:
\begin{equation}
H_c = \frac{\partial \ln (N\bar{\chi}/\tau_c)}{\partial \ln p}\approx H(1+\frac{\partial \ln\bar{\chi}}{\partial \ln p}-\frac{\partial \ln\tau_c}{\partial \ln p})
\end{equation}
where we have approximated the number density scale height using the pressure scale height $H$. In most cases, a chemical tracer does not quench to a vertically constant profile right at the departure point. We can also approximate the actual scale height of the tracer at around the departure point using the scale height of its equilibrium mixing ratio $\chi_{eq}$. Substitute $\bar{\chi}$ with $\chi_{eq}$ in Eq. (16) and the chemical scale height $H_c$ becomes the ``equilibrium chemical scale height" $H_{ceq}$ (\citealt{bordwell2018convective}). 

Using $L_v \sim H_{ceq}$, the ``effective dynamical timescale" (i.e., the right hand side of Eq. 15) is species-dependent (Fig. 6). Because the equilibrium mixing ratio $\chi_{eq}$ is the same for all species, the species with a steeper vertical gradient of $\tau_c$ (also the longer-lived species in our cases) has a smaller effective dynamical timescale. In our simulations, as pressure decreases, the equilibrium mixing ratio $\chi_{eq}$ decreases and chemical timescale $\tau_c$ increases. Therefore chemical destruction rate decreases faster than the pressure (or density) decrease, and thus $H_{eq}$ is smaller than $H$. This implies a smaller effective dynamical timescale, or a more efficient vertical transport, than the case using $L_v \sim H$. The predicted quench points are thus located in a relatively deeper atmosphere (Fig. 6 and Fig. 9). Overall, using $L_v \sim H_{ceq}$ could explain the departure points for the short-lived species but overestimates the pressure of quenching for long-lived tracers. In the latter regime, using $L_v \sim H$ gives a better prediction. 

\subsection{Results: Photochemical Experiment}
\begin{figure*}
\includegraphics[width=0.95\textwidth]{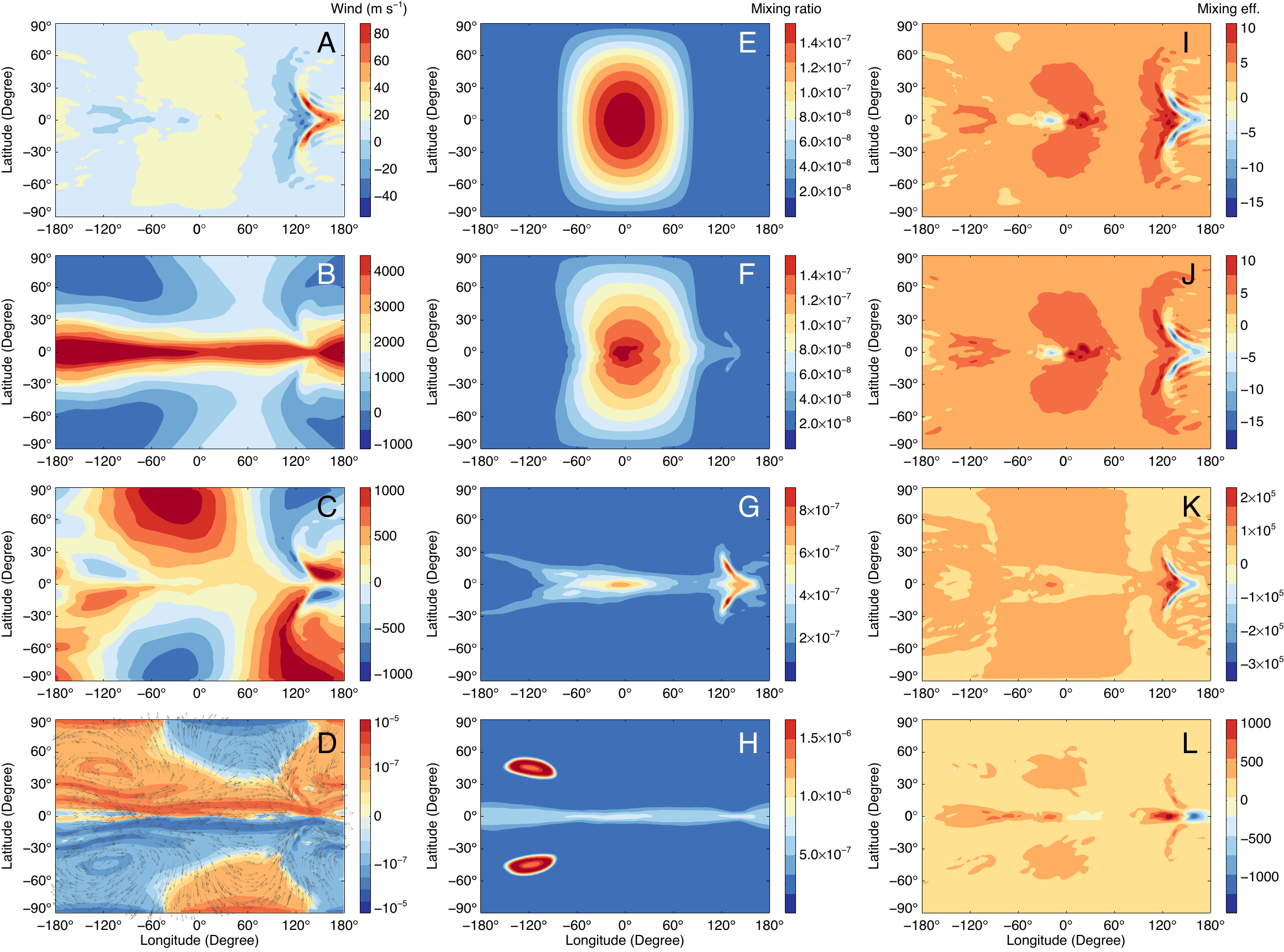} 
 \centering\caption{Longitude-latitude maps of various quantities at 2800 Pa in the Photochemical Experiment. Left column from top to the bottom (Panel A-D): vertical wind ($w$), east-west wind ($u$), north-south wind ($v$), and relative vorticity ($\zeta$) with wind field vectors. Central column (E-H): volume mixing ratios of tracers with chemical timescale of $10^{2.5}$ s, $10^{4}$ s, $10^{5.5}$ s, $10^{7} $s from top to the bottom, respectively. Right column (I-L): corresponding vertical mixing efficiency of the tracers in the central column. Note that the color bars do not have the same range.} 
\end{figure*}

Using the same dynamical field, we also investigated models where the tracer chemical equilibrium abundance varies strongly from the dayside to the nightside (Eq. 11). With a dayside chemical source, we can approximately simulate photochemically or ion-chemically produced species on tidally locked planets. The wind field and circulation have been discussed in the Standard case (Section 3.1). The tracer distributions look significantly different from previous cases (Fig. 10). For short-lived species, the tracer distribution is close to the prescribed chemical equilibrium distribution (Eq. 11) which maximizes at the sub-stellar point (Fig. 10E). The tracer abundance decreases away from the sub-stellar point and becomes approximately homogeneous on the nightside with a mixing ratio of $10^{-12}$. As the horizontal dynamical transport might be neglected when the tracer lifetime is short ($\tau_c\ll\tau_d$), the tracer distribution in this limit can be approximated based on Eq. (11) in Paper I:
\begin{equation}
\chi^\prime=\chi_0^{\prime}-w\tau_c\frac{\partial\overline{\chi}}{\partial z}.
\end{equation}
In this case, it is the deviation of the tracer mixing ratio from its chemical equilibrium distribution, instead of the tracer mixing ratio itself, that is anti-correlated with the vertical velocity pattern (Fig. 10A).

As the tracer chemical lifetime increases, atmospheric dynamics plays a more important role in shaping the tracer spatial distribution. The spatial pattern exhibits a ``chemical hot spot shift" on the dayside where the maximum abundance of tracer occurs slightly east of the sub-stellar point (Fig. 10F). This hot spot shift pattern resembles the well-known ``temperature hot spot" on tidally locked planets (e.g., \citealt{showman-guillot-2002}, \citealt{knutson2007map}). The chemical hot spot phase shift might also be detected by the light curve at the wavelengths where the species has significantly strong absorption or emission features. Physically, this chemical hot spot shift is a result of competition between the horizontal advection and chemical relaxation. Thus the phase shift could be predicted in a simple chemical-transport model in the longitudinal direction (\citealt{zhang2013jovian}) with a sinusoidal chemical equilibrium distribution. That theory states that the chemical hot spot phase shift $\phi_s$ is determined by the ratio of the chemical relaxation timescale to the advection timescale: $\phi_s=\tan^{-1}(\tau_cU/a)$. For a tracer with lifetime $\tau_c=10^4$ s (Fig. 10F), if we adopt the zonal-mean zonal wind velocity scale $U\sim2 \mathrm{~km~s^{-1}}$ at 2800 Pa (Fig. 10B), we predict a chemical hot spot phase shift $\phi_s\sim10$ degrees, roughly consistent with the simulation result. A more rigorous derivation to the phase shift can adopt the approach in the Appendix B in \citet{zhang2017effects} by replacing the temperature variable and radiative timescale with the chemical mixing ratio and chemical lifetime, respectively. But the predicted phase shifts from the above two approaches are generally very close. 

\begin{figure}[t]
\includegraphics[width=0.48\textwidth]{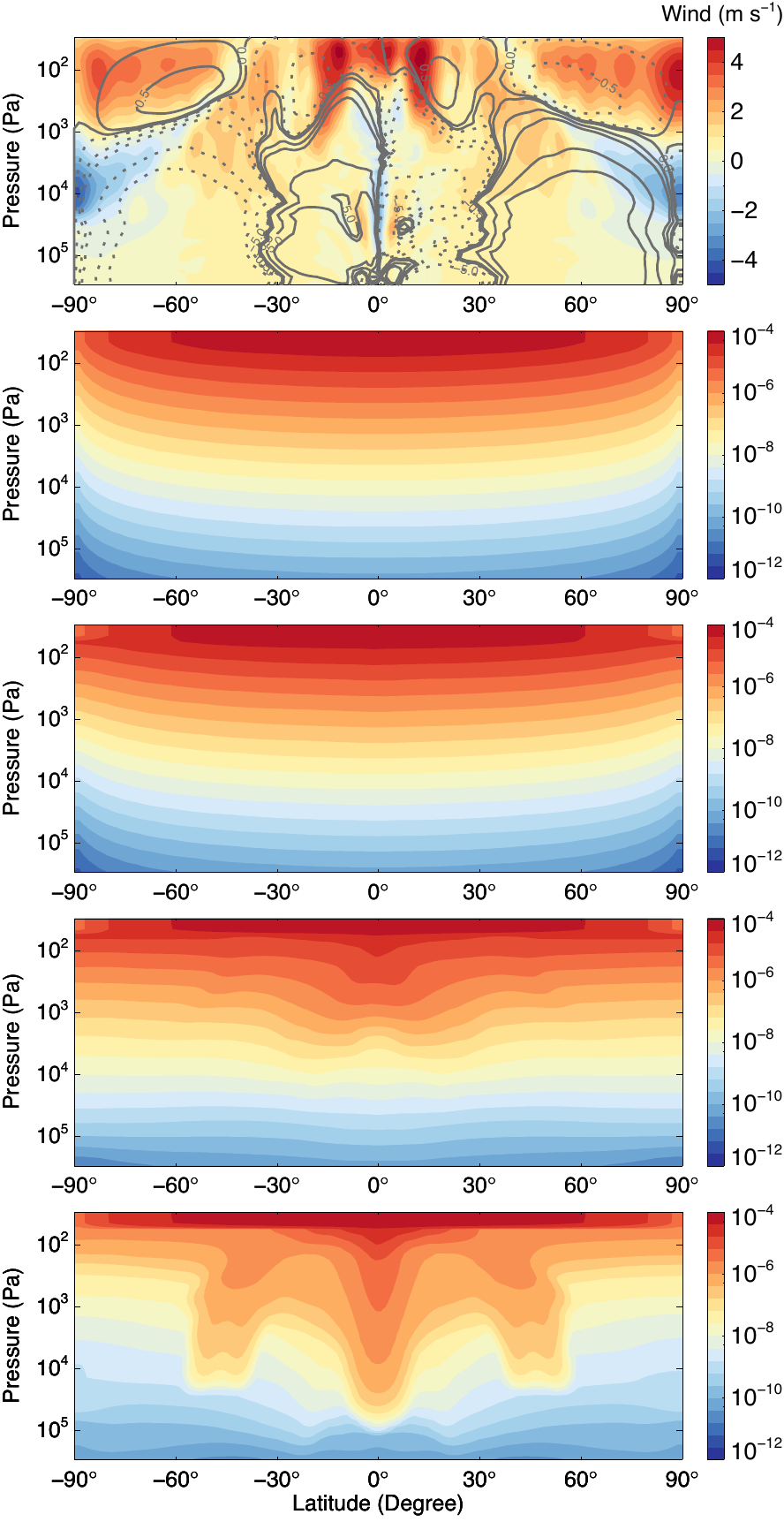} 
 \centering\caption{Latitude-pressure maps of zonally averaged quantities in the Photochemical Experiment. Top: vertical wind velocity (color) and TEM mass streamfunction (contours, in units of $10^{12}$ Kg $\mathrm{s^{-1}}$, solid/positive is clockwise). Starting from the second row we show volume mixing ratio of tracers with different chemical timescales of $10^2$ s, $10^{3.5}$ s, $10^5$ s, $10^{6.5}$ s s from top to bottom, respectively.} 
\end{figure}

\begin{figure}[t]
\includegraphics[width=0.48\textwidth]{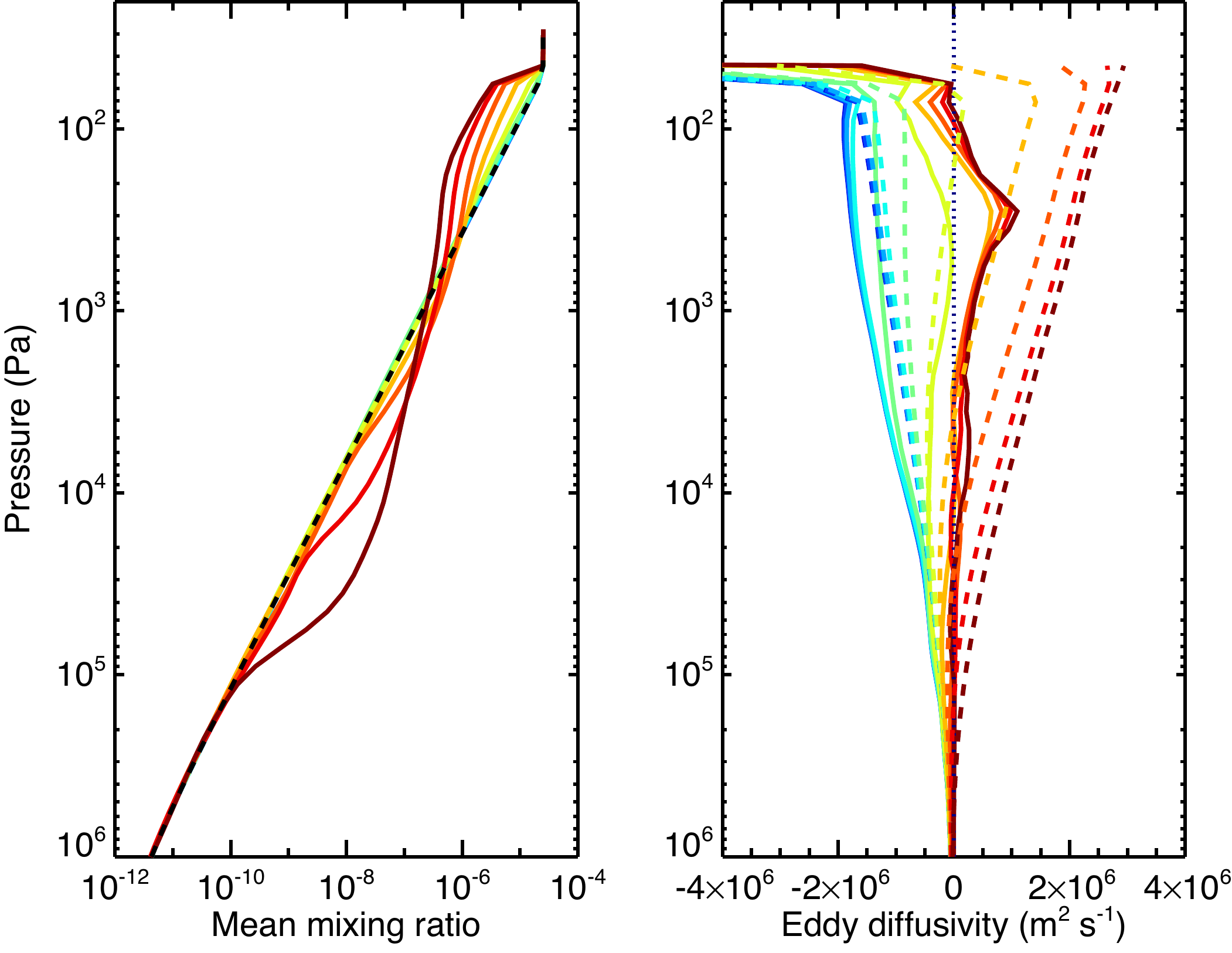} 
 \centering\caption{Left: vertical profiles of the global-mean volume mixing ratio in the Photochemical Experiment. Right: vertical profiles of numerically derived effective eddy diffusivity $K_{zz}$ (right) from the flux-gradient relationship (Eq. 8) based on simulation results (solid). The dashed lines are analytical predictions using Eq. (2) and the analytical $\hat{w}$ in Eq. (7). Different colors from cold (blue) to warm (red) represent tracers with different chemical timescales from short to long, ranging from $10^2$ s to $10^7$ s.} 
\end{figure}

If the tracer lifetime is significantly longer than both the east-west and north-south advection timescales, the spatial pattern of the tracer (Fig. 10G) is different from the``chemical hot spot" pattern. In our simulations, the tracers are more concentrated at low latitudes with some positive anomaly features near both terminator regions where local upwelling and downwelling are prominent in the vertical velocity map (Fig. 10A). The very long-lived tracer ($\tau_c=10^7$ s, Fig. 10H) is more homogeneously distributed compared with the short-lived tracers. But it exhibits two significant local anomalies, one in the northern hemisphere and the other in the southern, located between latitude 30 and 60 degrees and near longitude of west 120 degrees. The anomalies show tracer accumulations with respect to the surrounding background (Fig. 10H). Those two features might be a response to the in-situ large local wind gyres, which can be seen in the vorticity map (Fig. 10D). 

The mixing efficiency in the case with non-uniform chemical equilibrium abundance is fundamentally different from the case with uniform equilibrium abundance in Section 3.1. The mixing efficiency of the short-lived tracers is positive everywhere on the globe (Fig. 2I), implying that the tracers are all vertically mixed in the same direction. Negative mixing efficiency appears only in the long-lived tracer regime (e.g., Fig. 2L). However, in the non-uniform chemical equilibrium abundance case, even for the very short-lived species, mixing efficiency pattern shows large negative values in the upwelling region on the nightside (Fig. 10I-10L). This suggests that the tracers are not vertically mixed in the same direction everywhere and the global-mean vertical tracer transport is only a net difference between the upward and downward fluxes. The reason is that chemical equilibrium, instead of atmospheric dynamics, controls the spatial distribution of those short-lived chemical species. On the dayside, both the vertical velocity and tracer abundance deviation from the mean, $\chi_0^\prime$, are positive, and therefore in general contribute to the upward mixing. However, on the nightside, $\chi_0^\prime$ is negative due to its low abundance relative to the global-mean value, anti-correlated with the positive vertical velocity. As a result, a negative mixing efficiency occurs in the upwelling region between 120 and 180 degrees in the equatorial region (Fig. 10I). Compared with the uniform chemical equilibrium abundance case (Fig. 2), those localized ``chimneys" turn to ``sinks" of the global-mean tracer abundance as the strong upwelling plumes transport abundant low-mixing-ratio tracers to the upper atmosphere. 

Zonally averaged tracer distributions (Fig. 11) illustrate how the upper atmospheric tracers are vertically mixed into the lower atmosphere. As shown in the bottom panel of Fig. 11, the long-lived tracers are not uniformly mixed downward by the circulation. Instead, the zonally averaged tracer pattern are significantly shaped by the circulation cells shown in the upper panel. Especially, at high latitudes above the 1000-Pa pressure level, the lower mixing ratio tracers from the lower atmosphere is actually circulated upward and transported to the lower latitudes by the two polar cells in the upper atmosphere. This transport leads to a decrease of the global-mean tracer mixing ratio in the upper atmosphere compared to the chemical equilibrium profile, as shown in the area-weight-mean tracer profile in Fig. 11. Only at pressure levels below 1000 Pa, the global-mean tracer mixing ratio of those long-lived species becomes larger than the chemical equilibrium value. Therefore the mean tracer profile is significantly controlled by the details of the atmospheric dynamics, and there is no ``quench point" in our simulations with non-uniform chemical equilibrium abundance near the top\footnote{Note that in these cases chemical timescales are constant with pressure. Quenching occurs more readily when the tracer chemical lifetime varies with pressure, as we have showed in the Quench Experiment.}.

We found that the global-mean net vertical tracer flux of short-lived species is positive, implying that the tracer is mixed upward to the upper atmosphere. This upward transport mainly occurs in the broad upwelling region on the dayside where the tracer abundance also shows a positive anomaly (Fig. 10A and 10E). Note that in this case the global-mean tracer mixing ratio is larger in the upper atmosphere where the source is located. This counter-gradient transport suggests a negative effective eddy diffusivity. We numerically derived $K_{zz}$ based on the simulation and flux-gradient relationship (Fig. 8). It is indeed negative for short-lived species. It implies that, in the globally average sense, those tracers are mixed towards their source in the upper atmosphere, which is physically counter-intuitive in the traditional chemical-diffusion framework. However, this phenomena confirms our theory in Section 2. A significant non-diffusive component contributes to this effect, a result from the correlation between the velocity field and the equilibrium chemical distribution. If we assume the global-mean vertical profile of the tracer is not significantly different from the chemical equilibrium profile (Eq. 11), with the non-diffusive correction we introduced in Section 2, $K_{zz}$ might still be predicted from first principles based on Eq. (2):
\begin{equation}
K_{zz}\approx\frac{\hat{w}^2}{\hat{w}/H+\tau_{c}^{-1}}-\frac{\hat{w}\Delta\chi_{eq}}{1+\hat{w}\tau_{c}/H}(\frac{\partial\Delta\chi_{eq}}{\partial z})^{-1}.
\end{equation}
Here we have assumed that the nightside chemical equilibrium abundance $\chi_n$ is much smaller than $\Delta\chi_{eq}$. For very short-lived species ($\tau_c\rightarrow 0$), the first term on the right hand side of Eq. (18) might be neglected, leading to a non-diffusive term only:
\begin{equation}
K_{zz}\approx-\hat{w}\Delta\chi_{eq}(\frac{\partial\Delta\chi_{eq}}{\partial z})^{-1}.
\end{equation}

This expression physically explains the negative ``$K_{zz}$" in the short-lived tracer simulations, even though the vertical transport is not important for those species which are mainly controlled by the local chemistry. This non-diffusive effect on the global-mean tracer transport has not been considered in current 1D models in planetary atmospheres. 

Using the analytical prediction of $\hat{w}$ from Eq. (7), we predicted $K_{zz}$ based on Eq. (18). It qualitatively agrees with the numerically derived $K_{zz}$ for both short-lived species and long-lived species. Eq. (18) also shows that the non-diffusive effect becomes less important than the diffusive term as the chemical timescale increases. In our simulations, the numerically derived $K_{zz}$ increases with tracer lifetime and becomes positive for long-lived species (Fig. 12). But Eq. (18) overestimates the $K_{zz}$ by about a factor of 5 in the long-lifetime regime. As noted in Paper I, if the material surface of tracer is distorted significantly, our theory in Section 2 is not a good predictor of the effective eddy diffusivity. Also, the assumed vertical gradient of the mean-tracer in the non-diffusive correction (Eq. 18) is not a good approximation for the long-lived tracers because the tracer distributions, both vertical and horizontal, may substantially deviate from the chemical equilibrium distribution.

It should be noted that the non-diffusive component does not always decrease $K_{zz}$ under uniform chemical equilibrium. The sign of the non-diffusive contribution depends on the correlation between the vertical wind ($w$) and chemical equilibrium tracer distribution ($\chi_0^{\prime}$), as well as the vertical gradient of the global-mean tracer profile. For instance, if we consider the atmosphere region above the photochemical source layer on a tidally locked planet, chemical equilibrium abundance is large at the bottom of this region and smaller at the top, implying a negative vertical global-mean tracer gradient. In this case, the $w-\chi_0^{\prime}$ correlation is still positive because the photochemical equilibrium abundance is still larger on the dayside than the nightside. But because the vertical tracer gradient is negative, the non-diffusive component will enhance the total upward global-mean tracer transport efficiency and thus $K_{zz}$ (Eq. 18). Another example is vertical mixing of mineral cloud particles on hot Jupiter. Cloud microphysics predicts that there might be more titanium and silicate clouds forming at the west terminator region than on the nightside (e.g., \citealt{powell2018formation}) due to colder temperature at the west limb. The spatial correlation between the vertical wind and microphysical equilibrium cloud tracer is complex in this situation and the importance of the non-diffusive contribution to $K_{zz}$ has yet to be investigated. Moreover, cloud settling represents a qualitatively different type of source/sink than the simple linear chemical relaxation scheme that we have adopted in this paper. It is currently unclear whether such a source/sink due to particle settling would lead to diffusive behavior, even under a simplified scenario that ignores large, externally imposed day-night variations in cloud production.

The complicated behavior of $K_{zz}$ in the case with non-uniform chemical equilibrium abundance, especially the negative $K_{zz}$ owing to non-diffusive effects, implies that it might be difficult to achieve a satisfactory understanding---or even a self-consistent model---of photochemical tracer transport on tidally locked exoplanets in a simple 1D framework with the same eddy diffusivity for all species. A 3D dynamical-chemical model, which might be computationally expensive, will be of a great help in revealing the underlying transport mechanisms and interactions between the chemical and dynamical processes.

\section{Conclusion and Discussion}

In this study we investigated tracer transport on tidally locked planets using a simple chemical source/sink scheme. Using the analytical vertical velocity theory from \citet{komacek2016atmospheric} and \citet{zhang2017effects}, we constructed a first-principles theory for the 1D eddy diffusivity $K_{zz}$ on this type of planet. We performed 3D tracer simulations using a GCM and derived the $K_{zz}$ from the globally averaged vertical transport flux from the model. We showed that the $K_{zz}$ derived from the 3D simulations agree with our analytical predictions. Therefore this study can serve as a theoretical foundation for future work on estimating or understanding the effective eddy diffusivity for global-mean vertical tracer transport on tidally locked exoplanets and slow-rotating planets with strong day-night contrast such as Venus.

Our detailed analytical and numerical analysis for 3D tidally locked planets basically agree with the conclusions we have drawn in Paper I (1-8 in the Conclusion Section) for fast-rotating planets. We also confirm that all three typical regimes in our $K_{zz}$ theory exist on tidally locked planets. Since this is the first time to validate these conclusions in a 3D model with a strongly 3D circulation in a tidally locked configuration, we briefly recap all the points here. 

(1) Larger vertical velocities lead to a larger global-mean vertical tracer mixing. (2) Efficient horizontal eddy mixing reduces the horizontal variations of tracer and decreases the global-mean eddy diffusivity. (3) Global-mean eddy diffusivity depends on the tracer sources and sinks due to chemistry and microphysics. The effective eddy diffusivity increases with the chemical lifetime in the case with linear chemical relaxation. (4) In regime I, short-lived species exhibit a similar spatial pattern as the vertical velocity field. But the resulting tracer distribution is complicated in other regimes. (5) The diffusive assumption is generally valid in regime I and the effective eddy diffusivity is always positive using our idealized chemical schemes. But non-diffusive effects can be important in regime II if there is a good correlation between the equilibrium tracer field and the vertical velocity field. (6) Non-diffusive behavior could occur in regime III when the tracer chemical lifetime is much longer than the atmospheric dynamical timescale. (7) We derived the analytical species-dependent eddy diffusivity for tracers for the tracers on tidally locked planets, the theoretical predictions generally agree with our 3D simulations in all regimes. (8) In the 3D planetary atmospheres under tidally locked configuration, the widely used assumption in current 1-D chemistry and cloud formation models---a single profile of vertical eddy diffusivity for all species---is generally invalid. 

In this study, based on the species-dependent eddy diffusivity theory, we also provide a new analytical theory of the global-mean departure ``quench" point of the tracer from its chemical equilibrium profile. Because the species-dependent eddy diffusivity is smaller than the conventional species-independent value, the departure point estimated in our theory is located at lower pressure than the previous estimate if the tracer chemical timescale increases with decreasing pressure. For tracers on close-in tidally locked exoplanets, the departure point estimated from our theory assuming $L_v\sim H$ agrees with the numerical simulations for longer-lived species. For shorter-lived species, using the equilibrium chemical scale height $H_{ceq}$ as $L_v$ provides a better estimate.

How to better approximate the 3D tracer transport in a 1D model for real atmospheres? We emphasize that it is the correlation between the horizontally varying tracer fields and horizontally varying vertical velocity field along an isobar that determines the the vertical transport efficiency. Here we propose two approaches. First, from a theoretical point of view, it is necessary to perform 3D tracer transport simulations with more realistic chemical and microphysical schemes for a specific atmospheric situation (e.g., \citealt{lee2016dynamic}, \citealt{lines2018simulating}) and analyze how to better parameterize the 1D effective tracer transport in that atmospheric regime. Taking close-in tidally locked planets as an example, one could include simplified dayside photochemistry and ion chemistry to study a chemically active species, and/or include nightside haze/cloud condensation and dayside particle evaporation to study cloud particle transport. \citet{parmentier20133d} and this study represent an initial step in this direction. 

Second, from an observational point of view, it would be interesting to perform a correlation analysis between the observed tracer fields and the vertical velocity field (or the simulated vertical velocity field if the vertical velocity is not easy to determined from observations). Even for exoplanets, observations are starting to reveal horizontal gas distributions (\citealt{stevenson2017spitzer}) and the cloud distribution from the light curve data (e.g., \citealt{parmentier2016transitions}). A correlation analysis between the tracer fields and the velocity field estimated from the atmospheric models (e.g., \citealt{dobbs-dixon-lin-2008}; \citealt{showman-etal-2009}; \citealt{rauscher-menou-2010}; \citealt{heng-etal-2011}; \citealt{perna-etal-2010}; \citealt{mayne2014unified}) may shed light on how the vertical tracer transport operates on those planets and lead to a better understanding of global-mean vertical eddy mixing for future chemical models and cloud models. 

\section{Acknowledgements}		
This research was supported by NASA Solar System Workings Grant NNX16AG08G to X.Z. and A.P.S.. We dedicate this work to Dr. Mark Allen (1949-2016), one of the founders of the Caltech/JPL kinetics model. We thank T. Komacek, V. Parmentier and X. Tan for helpful discussions. Some of the simulations were performed on the Stampede supercomputer at TACC through an allocation by XSEDE.

\bibliographystyle{apj}
%\bibliography{../../../ZTexPaper/Reflib/Reflib.bib}

\end{document}